\documentclass[11pt]{article}%
\usepackage{amsmath}
\usepackage{amsfonts}
\usepackage{amssymb}
\usepackage{graphicx}%
\usepackage{algpseudocode}
\usepackage{lscape}
\usepackage{algorithmicx}
\usepackage[dvipsnames]{xcolor} %\usepackage{xcolor}
\usepackage{mathtools}
\usepackage{changes}
\usepackage{bbm}
\usepackage{ulem}

\usepackage[ruled, linesnumbered, vlined]{algorithm2e}
\usepackage{float}
\usepackage{tabularx}
\usepackage{natbib}
\usepackage{appendix}

\providecommand{\U}[1]{\protect\rule{.1in}{.1in}}
\newtheorem{theorem}{Theorem}

\newtheorem{definition}[theorem]{Definition}

\newtheorem{proposition}[theorem]{Proposition}

\newtheorem{fact}[theorem]{Fact}
\newenvironment{proof}[1][Proof]{\noindent\textbf{#1.} }{\ \rule{0.5em}{0.5em}}

\newcommand{\TREE}[1][A]{\ensuremath{\mathcal{#1}}} % ALBERO: \TREE \TREE[A]
\newcommand{\TREET}{\TREE[T]} 
\newcommand{\TREEA}{\TREE[A]} 
\newcommand{\TREEB}{\TREE[B]} 
 
\newcommand{\SUBTREE}[2][i]{\ensuremath{\TREE[{#2}]^{({#1})}}} % I-mo SOTTOALBERO: \SUBTREE{A} \SUBTREE[i]{A}

\newcommand{\SET}[1][A]{\ensuremath{\mathbf{#1}}}   % INSIEME DEI NODI DELL'ALBERO: \SET \SET[A]
\newcommand{\SETT}[1][T]{\ensuremath{\mathbf{#1}}}   % INSIEME DEI NODI DELL'ALBERO PRINCIPALE
\newcommand{\SUBSET}[2][i]{\ensuremath{\SET[{#2}]^{({#1})}}} % NODI I-mo SOTTOALBERO: \SUBSET{A} \SUBSET[i]{A}

\newcommand{\ROOT}[1][A]{\ensuremath{\textbf{\lowercase{#1}}}} % RADICE DELL'ALBERO \ROOT \ROOT[A]
\newcommand{\SUBROOT}[2][i]{\ensuremath{\ROOT[{#2}]^{({#1})}}} % NODO RADICE I-mo SOTTOALBERO: \ROOT{A} \ROOT[i]{A}

\newcommand{\ARC}[1][i]{\ensuremath{\textit{\lowercase{e}}^{({#1})}}} % 
\newcommand{\DIST}[1]{\ensuremath{d_{#1}}} % DISTANZA DA RADICE ALBERO: \DIST{\TREE{A}} \DIST{\SUBSTREE[i]{A}
\newcommand{\DISTA}{\DIST{\ROOT[A]}}
\newcommand{\DISTB}{\DIST{\ROOT[B]}}
\newcommand{\DISTT}{\DIST{\ROOT[T]}}

\newcommand{\PNODE}[1]{\ensuremath{\pi_{\lowercase{#1}}}} % PROBABILITA' DI UN NODO
\newcommand{\PROOT}[1][A]{\PNODE{\ROOT[#1]}} % PROBABILITA' DEL NODO RADICE
\newcommand{\PROOTA}{\PROOT[A]}

\newcommand{\PRIZE}[1]{\textbf{\ensuremath{p_{\lowercase{#1}}}}} % PREMIO DEL NODO RADICE

\newcommand{\PSETS}[1]{\ensuremath{\Pi_{#1}}} % PROBABILITA' DI UN INSIEME DI NODI: \PSETS{\SET[A]}
\newcommand{\PCOMP}[1]{\ensuremath{1-\PSETS{#1}}}

\newcommand{\PNOSETS}[1]{\ensuremath{\prod_{s\in {#1}}\left(1-\PNODE{s}\right)}} % PROBABILITA' CHE NESSUN NODO DI UN INSIEME EMETTA RICHIESTA

\newcommand{\EVENTO}{\ensuremath{\omega}}
\newcommand{\INDICATRICE}[2]{\ensuremath{\mathbbm{1}_{#1}(#2)}}
\newcommand{\VALOREATTESO}[1]{\ensuremath{\mathbb{E}\left[#1\right]}}

\newcommand{\RRICAVO}[1]{\ensuremath{R_{\EVENTO}(#1)}}
\newcommand{\MRICAVO}[1]{{{\ensuremath{\textbf{R}(#1)}}}}

\newcommand{\RCOSTO}[2]{\ensuremath{C_{\EVENTO}(#1;#2)}}
\newcommand{\MCOSTOB}[2]{\ensuremath{\textbf{C}(#1;#2)}}

\newcommand{\RGUADAGNO}[2]{\ensuremath{G_{\EVENTO}(#1;#2)}}
\newcommand{\MGUADAGNOB}[2]{\ensuremath{\textbf{G}(#1;#2)}}

\newcommand{\VBENZ}[2][x]{\ensuremath{\DISTA\PSETS{#2}+{#1}(\PCOMP{#2})}}

\newcommand{\TREEINDEX}[1][A]{\ensuremath{N_{\TREE[#1]}}}

\newcommand{\SETINDEX}[2]{\ensuremath{\SET[{#1}]_{#2}}}
\newcommand{\SETAINDEXED}[1][I]{\ensuremath{\SETINDEX{A}{#1}}}
\newcommand{\SUMTREE}{{\ensuremath{\sum_{i=0}^{m}}}}

\newcommand{\FC}[2][x]{\ensuremath{f^{({#2})}({#1})}}
\newcommand{\FCA}[1][x]{\FC[#1]{\TREEA}}

\newcommand{\FCSA}[1][x]{\FC[#1]{\SUBTREE{A}}}

\newcommand{\FCSETINDEXA}[2][x]{\FC[#1]{\TREE[A];{#2}}}

\newcommand{\DESCRIZIONE}[1]{\ensuremath{\mathcal{D}_{#1}}}
\newcommand{\DESCRIZIONEA}{\DESCRIZIONE{\TREEA}}
\newcommand{\DESCRIZIONEAI}[1][I]{\DESCRIZIONE{(\TREEA;#1)}}
\newcommand{\DESCRIZIONEAB}{\DESCRIZIONEAI[\GRUPPOB]}
\newcommand{\DESCRIZIONEAC}{\DESCRIZIONEAI[\GRUPPOC]}

\newcommand{\DESCRIZIONEAUBC}{\DESCRIZIONEAI[\GRUPPOUBC]}

\newcommand{\MATSET}[2]{\ensuremath{\texttt{#1}_{#2}}}

\newcommand{\MATSETB}[1]{\MATSET{B}{#1}}
\newcommand{\MATSETC}[1]{\MATSET{C}{#1}}
\newcommand{\MATSETD}[1]{\MATSET{D}{#1}}
\newcommand{\MATSETK}[1]{\MATSET{K}{#1}}

\newcommand{\GRUPPO}[1]{\ensuremath{#1}}
\newcommand{\GRUPPOB}{\GRUPPO{B}}
\newcommand{\GRUPPOC}{\ensuremath{C}}
\newcommand{\GRUPPOD}{\ensuremath{D}}
\newcommand{\GRUPPOUBC}{\ensuremath{\GRUPPOB\cup\GRUPPOC}} %\newcommand{\GRUPPOUBC}{\ensuremath{\mathbbm{D}}}

\newcommand{\MATRIOSKA}[1]{\ensuremath{\mathcal{M}_{#1}}}
\newcommand{\MATRIOSKAA}{\MATRIOSKA{\TREEA}}
\newcommand{\MATRIOSKAAI}[1][I]{\MATRIOSKA{(\TREEA;#1)}}

\newcommand{\MATRIOSKAB}{\MATRIOSKAAI[\GRUPPOB]}
\newcommand{\MATRIOSKAC}{\MATRIOSKAAI[\GRUPPOC]}
\newcommand{\MATRIOSKAD}{\MATRIOSKAAI[\GRUPPOD]}
\newcommand{\MATRIOSKAUBC}{\MATRIOSKAAI[\GRUPPOUBC]}

\newcommand{\MATRIOSKAK}{\MATRIOSKA{K}}
\newcommand{\MATRIOSKAKH}[1][h]{\ensuremath{\mathcal{M}_{K}^{(#1)}}}

\newcommand{\CARTMATRIOSKABIN}{\ensuremath{\MATRIOSKAB\times\MATRIOSKAC}}
\newcommand{\CARTMATRIOSKA} {\ensuremath{\MATRIOSKA{\SUBTREE[0]{A}}\times\MATRIOSKA{\SUBTREE[1]{A}}\times\MATRIOSKA{\SUBTREE[2]{A}}\times\ldots\times\MATRIOSKA{\SUBTREE[m-1]{A}}\times\MATRIOSKA{\SUBTREE[m]{A}}}}

\newcommand{\CARDA}{n_A}
\newcommand{\CARDB}{n_B}
\newcommand{\CARDC}{n_C}
\newcommand{\CARDD}{n_D}

\newcommand{\ENVDIS}[3]{\ensuremath{\DESCRIZIONEAI[#1]^{(#2)}.{#3}}}

\newcommand{\FIELDSET}{\ensuremath{S}}
\newcommand{\FIELDXMIN}{\ensuremath{x^{\min}}}
\newcommand{\FIELDXMAX}{\ensuremath{x^{\max}}}
\newcommand{\FIELDPENDENZA}{\ensuremath{\pi}}
\newcommand{\FIELDQZERO}{\ensuremath{q}}

\newcommand{\RDESCRIZIONEI}[3]{\ensuremath{\mathcal{R}_{(\TREE[{#2}];#3)}^{(#1)}}} % LOCALE
\newcommand{\RDESCRIZIONEAI}[2][I]{\RDESCRIZIONEI{#2}{A}{#1}}

\newcommand{\RDXMINBC}[2]{\ensuremath{\RDESCRIZIONEI{#2}{A}{#1}.\FIELDXMIN}}
\newcommand{\RDXMAXBC}[2]{\ensuremath{\RDESCRIZIONEI{#2}{A}{#1}.\FIELDXMAX}}
\newcommand{\RDPENDENZABC}[2]{\ensuremath{\RDESCRIZIONEI{#2}{A}{#1}.\FIELDPENDENZA}}

\newcommand{\RDSET}[2]{\ensuremath{\RDESCRIZIONEI{{#2}}{#1}{I}.\FIELDSET}} % LOCALE
\newcommand{\RDXMIN}[2]{\ensuremath{\RDESCRIZIONEI{{#2}}{#1}{I}.\FIELDXMIN}} % LOCALE
\newcommand{\RDXMAX}[2]{\ensuremath{\RDESCRIZIONEI{{#2}}{#1}{I}.\FIELDXMAX}} % LOCALE
\newcommand{\RDPENDENZA}[2]{\ensuremath{\RDESCRIZIONEI{{#2}}{#1}{I}.\FIELDPENDENZA}} % LOCALE
\newcommand{\RDASETI}[1]{\RDSET{A}{#1}}
\newcommand{\RDAXMINI}[1]{\RDXMIN{A}{#1}}
\newcommand{\RDAXMAXI}[1]{\RDXMAX{A}{#1}}
\newcommand{\RDAPENDENZAI}[1]{\RDPENDENZA{A}{#1}}

\newcommand{\RDESCRIZIONE}[2]{\ensuremath{\mathcal{R}_{\TREE[{#2}]}^{(#1)}}} % LOCALE
\newcommand{\RDESCRIZIONEA}[1]{\RDESCRIZIONE{#1}{A}}

\newcommand{\RDTSET}[1]{\RDESCRIZIONE{#1}{T}.\FIELDSET}
\newcommand{\RDTQZERO}[1]{\RDESCRIZIONE{#1}{T}.\FIELDQZERO}
\newcommand{\RDTPENDENZA}[1]{\RDESCRIZIONE{#1}{T}.\FIELDPENDENZA}

\newcommand{\SOGLIA}[2]{\ensuremath{\mathcal{S}(#1;#2)}}
\newcommand{\REQUIREREC}[3]{\ensuremath{\frac{\RDXMINBC{#1}{#2}-\DISTA\PSETS{#3}}{\PCOMP{#3}}}}
\newcommand{\REQUIRERECMAX}[3]{\ensuremath{\frac{\RDXMAXBC{#1}{#2}-\DISTA\PSETS{#3}}{\PCOMP{#3}}}}

\newcommand{\PROFITLINE}{{{set profit line}}}
\newcommand{\PROFITLINES}{{{set profit lines}}}
\newcommand{\TREECHARFUNCTION}{{{tree characteristic function}}}
\newcommand{\SUBTREECHARFUNCTION}{{{sub-trees characteristic function}}}
\newcommand{\SUBTREECHARFUNCTIONS}{{{sub-trees characteristic functions}}}
\newcommand{\SETCHARFUNCTION}{\PROFITLINE} %{\red{\textbf{set characteristic function}}}
\newcommand{\SETCHARFUNCTIONS}{\PROFITLINES}

\newcommand{\OPTIMALSET}{{{optimal set}}}
\newcommand{\OPTIMALSETS}{{{optimal sets}}}
\newcommand{\MAXOPTIMALSET}{{{maximal optimal set}}}
\newcommand{\MAXOPTIMALSETS}{{{maximal optimal sets}}}
\newcommand{\MINOPTIMALSET}{{{minimal optimal set}}}

\newcommand{\SOGLIAINGRESSO}{{entry threshold}}
\newcommand{\SOGLIEINGRESSO}{{entry thresholds}}

\title{Solving the Probabilistic Profitable Tour Problem on a tree}
\author{Enrico Angelelli$^{(1)}$, Renata Mansini$^{(2)}$, Romeo Rizzi$^{(3)}$\\ \\
	{\small $(1)$  Department of Economics and Management, University of Brescia, Italy,
		{\tt enrico.angelelli@unibs.it}} \\
	{\small $(2)$  Department of Information Engineering, University of Brescia, Italy,
		{\tt renata.mansini@unibs.it}}\\
	{\small $(3)$ {Department of Computer Science, University of Verona, Italy },
		{\tt romeo.rizzi@univr.it}}\\
}
\date{}

\begin{document}
	
	\maketitle
	\begin{abstract}
% 	Probabilistic problems represent important generalization of combinatorial problems that  deal with uncertainty through the  inclusion of some probabilistic aspects.  
	
% 	In this paper, we analyze 
% 	the probabilistic variant of the profitable tour problem (PTP)  a well-known NP-hard routing problem searching for  a tour visiting a subset of
% customers while maximizing profit as difference between total revenue collected from visited customers
% and traveling costs to visit them. 
% Although PTP is solvable in
% polynomial time when special structure of the underlying graph are taken into account, the computational complexity of the corresponding probabilistic variants are still open issues.
% In this paper, we analyze the probabilistic PTP 
% where customers are located on a tree and need, with a known probability,  for a service provision  for which they pay a predefined prize. A business contract has to be established in advance with a subset of such customers before knowing if they will ask or not to be serviced. The problem objective is to maximize the expected profit obtained by visiting  only the customers that require a service among the ones with a contract. 
The profitable tour problem (PTP) is a well-known NP-hard routing problem searching for a tour visiting a subset of customers while maximizing profit as the difference between total revenue collected and traveling costs. 
PTP is known to be solvable in polynomial time when special structures of the underlying graph are considered. However, the computational complexity of the corresponding probabilistic generalizations is still an open issue in many cases.
In this paper, we analyze the probabilistic PTP where customers are located on a tree and need, with a known probability, for a service provision at a predefined prize. 
The problem objective is to select a priori a subset of customers with whom to commit the service so to maximize the expected profit. 
We provide a
polynomial time algorithm computing the optimal solution in $O(n^2)$, where $n$ is the number of nodes in the tree.
	\end{abstract}
	

\section{Introduction}
%\note{Stile citazioni nome+numero vs. solo numero.}
Routing problems with profits (prizes) %(revenues) 
represent some of the most challenging variants of vehicle routing problems. 
%The presence of revenues assigned to customers  and constraints on vehicles traveling times  imply that  not all customers have to be visited.
% provided that collected revenues are maximized. 
Several real-world routing problems jointly consider the optimization of  revenues %(collected when visiting customers) 
and transportation costs, 
%required to reach them, 
the latter either bounded in the constraints or directly subtracted from the global revenue in the objective function (\cite{FDG05}).

In the literature, a gamut of contributions can be found on deterministic  variants of single and multi vehicle routing problems combining {prize} collection to other real constraints 
(see, for instance, \cite{yu2022team} where arrival and
service times affect the collection of {profits}, \cite{HMZ20} where multi-visits are allowed while complying with precedence constraints and \cite{doi:10.1137/1.9781611973594.ch10} for a survey). 
Nevertheless, a limited number of papers deal with these routing problems under uncertainty.

Probabilistic problems,  characterized by the  inclusion of some probabilistic elements in the problem definition, can be seen as generalizations  of the corresponding deterministic variants. 
Different features can be probabilistic in a routing problem: from traveling times, arc costs, service times, to customer requests and collected prizes.
A large part of the literature has been focused on problems where  the probabilistic feature in the graph is associated with  nodes (see, for instance, %\cite{murat1999} and 
\cite{bellalouna1995} and \cite{henchiri2014} and references therein).
%\red{[]alcuni esempi e magari una SURVEY]}. % that  have  a specified probability to be present.
%We are interested in tackling combinatorial optimization problems on subsets of nodes by finding an optimal strategy so that the problem is optimally solved for every instance.
The introduction of probabilistic aspects %is 
typically  %two-fold. On one side, it 
allows the decision maker to face more realistic problems in which uncertainty is a major concern. It can be used as a tool to make 
 strategic planning, for instance to plan a service/product distribution by evaluating its expected convenience. 
Probabilistic problems are commonly tackled in literature by means of an {\it a  priori optimization} strategy. 
It consists in finding  an a priori solution 
 and then applying a modification
strategy to adapt this solution to the occurrence (\cite{BJO90}). It is evident how such a strategy can be expressed as a 2-stage decision scheme. 
For example, in \cite{laporte1994}, the a priori optimization is applied to  the  probabilistic traveling salesman problem.   
In a first stage, the computation of  an a priori Hamiltonian tour is obtained and then,  in a second stage, when information on the present nodes has been revealed, the a priori tour is followed by skipping the absent nodes (recourse action).
%Recourse action might require the solution of a problem of high computational complexity.  
%When this is not the case,  
%the resulting 2-stage problem is  easy to solve. This is the case of the problem analyzed in this paper.

In the literature different works can be found on the  probabilistic variants of both  traveling salesman problem and vehicle routing problem (see \cite{jaillet1985}, \cite{jaillet1988}, \cite{bertsimas1993} and \cite{campbell2008}). 
In case minimization of waiting time instead of distance traveled is considered, the problem becomes a repairman problem whose probabilistic variant on a tree has been studied by \cite{10.2307/25768684}.
In \cite{AAFV17} the authors analyze
the probabilistic orienteering problem 
providing a
stochastic mixed integer linear program and  deriving  a deterministic equivalent formulation to solve  by means of  an exact branch-and-cut method.

In this paper, we consider a probabilistic routing problem  where 
a company has to decide which potential customers to engage into a long-term business  agreement (typically an annual contract) for service provision (e.g. assistance on demand) so to maximize the expected value of its daily profits (revenues minus traveling costs).
More precisely, 
according to the agreement,   a customer will receive the service in a given day as long as he/she has requested it the day before. The service is guaranteed and provided by the company against the payment of a prize contractually established (revenue for the company), which is paid by the customer for each service call.
The company does not know how many times (if any) each potential customer will need the service and when during the year. 
However, it knows each customer, on every single day, will call for a service with a given probability. In practice,  the company will serve a (possibly) different set of customers every day.  The service will be carried out at the minimum operating cost measured as  travel cost (distance) required to visit in a tour all the customers involved.
The profit (revenue minus cost) earned daily by the company depends on the set of customers under contract and on those who, among them, issue a service request.
Therefore, the profit is a random variable that depends on the set of customers under contract, and each one of its  realizations (the daily profit) is related to the customers 
issuing a service request  on each day. Since the global profit over the year is given by the sum of daily realizations, the goal of the company is to determine with which  customers it will undertake a contract so to maximize the expected value of the daily profit. 
In terms of a 2-stage decision process, the company first selects the subset of potential customers entering the annual contract and for which the provided service will be guaranteed (a priori decision),  then, every day,  the service requests from customers occur (nature outcome), 
and as a recourse action the company serves the customers requiring the service at the  minimum {travel} cost.
This generalization of the deterministic profitable tour problem (PTP) is known as  the Probabilistic Profitable Tour Problem (PPTP).
In \cite{zhang2017}, the authors study this problem applied to the case of  
big e-tailing companies that need to determine, among the set of all potential customers, which customers to serve directly and which  outsource. %The problem is a probabilistic profitable tour
%looking for  a subset of customers  and an a priori tour through such customers  that maximizes the difference between the expected revenues and the expected travel costs. 

In our problem, the potential customers are assumed to be  geographically dispersed in a mountain area characterized by a main valley surrounded by a number of lateral valleys.  The {\it {road} network} consists of 
one main road that traverses the {main} valley down up and is characterized by the presence of several bifurcations, that in turns generate secondary roads, leading to the lateral valleys, with a similar structure to the main one. Each customer location is linked to the company's
depot from a unique path through the {road} network. 
%The corresponding routing network can be easily described as  a tree rooted at the depot where 
Thus, the {road} network  can be express as a {tree} rooted at the depot where the other nodes are customers or bifurcations traversed to reach farthest customers located on secondary roads. 
Edge weights  represent the traveling cost to traverse them. %twice (up and down).

%The analysis is now extended to the consideration of a tree problem.

When considering specific network structure (line, cycle, tree, star), the computational complexity of the deterministic PTP and  other deterministic variants of the traveling salesman problem with profits can be found in \cite{ABST14}. The specific case of deterministic PTP on trees was previously discussed in \cite{klau2003fractional} where a linear time algorithm is provided.
Finally, in \cite{AMR22}, the probabilistic PTP
is analyzed considering the special case of the underlying graph  represented by a path. 
In that paper, the authors show that the problem
can be solved in $O(n^2)$ time, where $n$ is the number of customers and provide an interesting  characterization of the optimal solution space of the problem. 

The present variant of the problem analyzed over  a rooted tree, from now on called the Probabilistic Profitable Tour Problem  on a Tree (PPTP-T),  is the generalization of the PPTP on  a path studied in \cite{AMR22}.  
%The corresponding  deterministic variant has been solved in polynomial time in ...
We show that also PPTP-T can be optimally solved in polynomial time with complexity $O(n^2)$
on a general rooted tree with $n$ nodes. 
%\remove{and no bounded vertex degree}. 

We believe that our study can be relevant also for other application domains where the tree-like  topology is commonly used (e.g. communication networks).

%Our analysis is restricted to a tree for different reasons. 
%The problem is obviously  very difficult for general  networks. Moreover, 
%Some 
%transportation (as the one described) and communication networks have tree or tree-like topology. Finally, 
%the analysis of the problem provides an insight to this specific topology and might help the  study of the problem on more general network.} 

The paper is organized as follows. 
Section \ref{SEC:notation} provides main notation and introduces the formal problem definition.  
In Section \ref{SEC:properties}, we discuss the main properties of 
%\replace{a \SUBTREECHARFUNCTION\ used to define \red{the dynamic programming} approach formulated to solve the problem.}
{the characteristic function of a generic sub-tree which we recursively compute on  larger sub-trees up to the main one.}
Finally, in Section \ref{SEC:algorithm}, we formalize the solution algorithm and discuss its computational complexity. 
Conclusions are  drawn in Section \ref{SEC:conclusions}.

\section{Notation and problem definition}\label{SEC:notation}
%
%The present section is devoted to discuss the formal definition of the PPTP-T.

The central contribution of this paper is a polynomial time algorithm for the Probabilistic Profitable Tour Problem (PPTP) when restricted on trees (PPTP-T).  
The present section is devoted to introduce the main notation used in the paper while providing a formal definition of both PPTP and PPTP-T.
In PPTP the topology is described by a general graph $\left<V,E\right>$ where 
node set $V$ includes customers and  the company's  depot, and
every edge $e\in E$ has a non-negative length that represents its traveling cost. 
%

%(in case of more customers displaced in the same location we can connect each one of their private avatar nodes to the location node, via an undirected edge of zero length). In this way every node of the graph is either a location node, or a customer node, or the depot node.
%
%
%\note{non si dice cosa facciamo se uin una location c'è un solo customer. Lo sdoppiamo o resta com'è? Nel secondo caso il nodo è entrambe le cose con una delle due. Nel primo caso avremo nodi con un solo discendente oppure foglie che non hanno distanza zero dal proprio genitore}
%
%
For every customer node $v\in V$, the input also specifies the probability $\pi_v>0$ that customer $v$ will actually issue a request and the prize $p_v>0$ collected by the company when fulfilling this request.
In fact, a Bernoulli random variable $r_v$ is associated with each customer node $v$ that
takes value $1$ with probability \PNODE{v} (node $v$ issues a request) and $0$ with probability $1-\PNODE{v}$ (no request).
Variables $r_v$ are assumed to be independent.
Thus, the probability that a subset $\EVENTO$ of the customers turns out to comprise precisely those customers willing to issue a request is $\mathbb{P}(\EVENTO)=\prod_{v\in\EVENTO}\PNODE{v}\cdot\prod_{v\notin\EVENTO}(1-\PNODE{v})$. 

Given all these data as its input, the company has to take an offline decision: select the most convenient subset $S$ of the customers with which to commit, knowing that, after this decision is taken, each customer $s$ in $S$ will issue its request (of value $p_s$) with probability $\pi_s$, independently one from the other (Bernoulli process). With the choice of $S$, the company has committed to fulfill each one of the issued requests in $S$, and is going to collect the relative prizes, but also to cover the whole costs of the cheapest tour that starts from the depot, visits each one of the customers in $S$ that have issued their request, and finally returns to the depot. 
%
%Denote with $\EVENTO_S$ the set of selected customers issuing their requests. Then $\EVENTO_S \subseteq S$ is a random variable whose distribution depends only on $S$ (and the $\pi_s,s\in S$ already specified by the given instance); in particular, for any set of customers $S'$ we have that
%\[
%   \mathbb{P}(\EVENTO_S=S')=\left\{
%       \begin{array}{ll}
%       \prod_{v\in S'}\PNODE{v}\cdot\prod_{v\notin S'}(1-\PNODE{v}) & \mbox{if $S' \subseteq S$,}\\
%       0 & \mbox{if $S' \not \subseteq S$.}\\
%       \end{array}
%    \right.
%\]

Given the a priori selection of $S$, we define and consider three random variables all depending on the outcome  $\EVENTO$: the \textit{revenue}, the \textit{cost}, and the \textit{profit} (\textit{revenue} - \textit{cost}).
The PPTP
asks to select $S$ with the objective to maximize the expected value of the profit.

%The revenue is $\sum_{s\in \EVENTO_S}\PRIZE{s}$ and its expected value finds a simple closed form expression in terms of $S$: $\sum_{s\in \EVENTO_S}\PRIZE{s}\pi_s$.

The  expected value of the revenue is easy to compute. On the contrary,  %Unfortunately, 
%\replace{for general graphs, even computing  the cost of a given outcome  $\EVENTO$ is NP-hard}
{computing  the cost of a given outcome  $\EVENTO$ is NP-hard on general graphs}
as it essentially amounts to the solution of a standard TSP  on node set $S\cap\EVENTO$. {Thus, }
 PPTP is an Np-hard problem. 
{Observe that, when facing PPTP, the outcome  $\EVENTO$ is not given, and $S$ is actually the independent variable we should optimize on.}

%
%
%\note{E' proprio necessario questo PIPPONE quando è chiaro che PTPP è una generalizzaizione di TSP?}
%
%

The PPTP-T is the restriction of PPTP on trees. Here, we assume the input graph to be an undirected 
%and unordered 
tree rooted at the depot.
Clearly, it is easy to identify a cheapest route to follow for a given $S$ and any outcome $\EVENTO$. As a matter of fact, every %\remove{feasible} 
solution to the resulting TSP, i.e., every optimal tour that starts from the root and ends in the root of the tree after having visited the node set $S\cap\EVENTO$, is 
limited to 
those tours in which every edge is traversed at most once in each direction. 
In particular, an edge should be considered if and only if it lays on the unique path from 
%\replace{a visited node}
{a node in $S\cap\EVENTO$}
to the root  of the tree.
For sake of simplicity, from now on, we assume that the cost of an edge includes both traversing cost
back and forth. 
%
%
%
%
%

%Without loss of generality we can assume that customers are located on the leaf nodes of the tree and internal nodes are only of bifurcation kind. 
%If this is not the case we can replace the customer internal node with a bifurcation node to which we append one more child node representing the customer at null distance from the bifurcation itself. 
%Even if we do this transformation on all internal nodes of a tree with $n$ nodes, we still have a tree with cardinality $O(n)$ nodes (i.e. no more than $2n$ nodes in all). 
%As bifurcation nodes are not customer to be selected, we can exclude them from any optimal solution by setting their prize and probability to $-1$ and $1$, respectively. 
%This way, there is no need to explicitly distinguish between client and bifurcation nodes in our analysis.
%
It might be  the case that some of the nodes the tree is built upon are not real customers, but just bifurcation nodes of the underlying road/communication network. 
On one side, we are not willing to select bifurcation nodes in our solution; on the other hand, for sake of notation and analysis simplicity, we prefer not to formally distinguish between customer and bifurcation nodes.
Thus, without any loss of generality, we assume that a prize equal to $-1$ and a probability equal to $1$ are assigned to each node representing just a tree bifurcation. %\color{red}
This is enough to guarantee that bifurcation nodes will be automatically excluded from any optimal node selection. % with no need to explicitly distinguish between client and bifurcation nodes in our analysis.
%\color{black}

\subsection{Notation}

%We now introduce the 
% main notation and formalize the network topology  %(the tree) 
%over which the PPTP-T is defined.

%
Let \TREET\ be a  \textit{rooted weighted tree} whose recursive structure is described on a generic sub-tree \TREEA\ as follows:
%
%A tree \TREEA\ consists of a root node and a list of $m\geq0$ sub-trees connected to the root:

$$\TREEA=\left\{
%(
\ROOT[A],
%\PROOTA,
%\PRIZE{\ROOT[A]},
%\DISTA),
\left[
%(\ARC[1]{A},\SUBTREE[1]{A}),
(\ARC[1],\SUBTREE[1]{A}),
\ldots,
%(\ARC[i]{A},\SUBTREE[i]{A}), \ldots,
%(\ARC[m]{A},\SUBTREE[m]{A})
(\ARC[i],\SUBTREE[i]{A}), \ldots,
(\ARC[m],\SUBTREE[m]{A})
\right]
\right\}
$$
where \ROOT[A] is the root of \TREEA, \SUBTREE[i]{A} is the $i$-th sub-tree of \TREEA,   and $\ARC[i]\geq 0$ is the {cost} required to reach its root \SUBROOT[i]{A} 
%of \SUBTREE[i]{A} 
starting from the root  \ROOT[A] of  \TREEA.
The cost required to reach the root \ROOT[A] from the root \ROOT[T] of the main tree \TREET\ is indicated with \DISTA, and we naturally have $\DISTT=0$ for the root \ROOT[T] of the main tree \TREET, and  $\DIST{\SUBROOT{A}}=\DISTA+\ARC[i]$ in general.
%$\PROOTA$ %$\in(0,1]$ 
%its  probability and 
%$\PRIZE{\ROOT[A]}$ %$\neq 0$ 
%the offered prize, whereas 
If 
$m=0$, the tree \TREEA\ consists only of the root node (leaf node of \TREET) and is indicated as
$\TREEA=\left\{
%(
\ROOT[A],
%\PROOTA,
%\PRIZE{\ROOT[A]},
%\DISTA),
\left[\ \right]
\right\}.
%\quad\text{with } m=0.
$

{Only for sake of notation simplicity}, 
when $m>0$ we indicate as \SUBTREE[0]{A} the root node intended as a sub-tree of a single node whose edge connection \ARC[0]  %\ARC[0]{A} 
measures zero. This will allow us to describe tree $\TREEA$ as follows: 
$$\TREEA=\left\{
\left[
%(\ARC[0]{A}=0,\SUBTREE[0]{A}),
%(\ARC[1]{A},\SUBTREE[1]{A}),
%\ldots,
%(\ARC[m]{A},\SUBTREE[m]{A})
(\ARC[0]=0,\SUBTREE[0]{A}),
(\ARC[1],\SUBTREE[1]{A}),
\ldots,
(\ARC[m],\SUBTREE[m]{A})
\right]
\right\}
\quad\text{with } \SUBTREE[0]{A}= \left\{
%(
\ROOT[A],
%\PROOTA,
%\PRIZE{\ROOT[A]},
%\DISTA),
\left[\ \right]
\right\}.
 %m>0.
$$
%\note{DOMANDA: Cosa vuoi far vedere come figura? Non mi sembra strettamente necessaria ma parliamone.... }
%
%
%\note{set of nodes -- node set: uniformare? (non mi sebra comunque un problema)}
%
%
We indicate as \SET\ the set of nodes contained into the tree  \TREEA\ and with  \SUBSET{A} the node set associated with the sub-tree \SUBTREE{A}, so that 
%$\SET = \{\ROOT\}\bigcup\left(\bigcup_{i=1}^{m}\SUBSET[i]{A}\right)$, or equivalently, 
$\SET = \bigcup_{i=0}^{m}\SUBSET[i]{A}$.
Defined as $\TREEINDEX=\{0,1,\ldots,m\}$ the index set of the sub-trees of \TREE, for each  $I\subseteq\TREEINDEX$ we denote as $\SETAINDEXED=\bigcup_{i\in I}\SUBSET{A}$ the set of nodes belonging to the group of sub-trees indexed in  $I$, so that $\SET = \SETAINDEXED[\TREEINDEX]=\bigcup_{i\in\TREEINDEX}\SUBSET[i]{A}$.

%A Bernoulli random variable $r_s$ is associated with each node $s\in\SETT$. $r_s$ takes value $1$ with probability \PNODE{s} (node $s$ issues a request of service), and $0$ with probability $1-\PNODE{s}$ (no request is issued by node $s$).
%
%Variables $\{r_s\ |\ s\in\SETT \}$ are assumed to be independent, thus, the probability that a request of service is issued by a subset of nodes $\EVENTO\subseteq\SETT$ is $\mathbb{P}(\EVENTO)=\prod_{v\in\EVENTO}\PNODE{v}\cdot\prod_{v\notin\EVENTO}(1-\PNODE{v})$.
%
%{\color{blue} From now on, we refer to \EVENTO\ as the set of nodes that ask for a service ({\it nature outcome}). }

%Each node $s\in\SETT$ is associated with  a non null prize \PRIZE{s} offered only in the case the node issues a request of service.
%

%\remove{The tree \TREET\  represents the road network containing both customers and bifurcations. 
%Thus, without loss of generality, to exclude bifurcation nodes $s$ from the a priori selection, we set 
%a negative prize (i.e. 
%$\PRIZE{s}=-1$ and 
%with probability 
%$\PNODE{s}=1$
%for all such nodes. % $s$ representing bifurcations.
%}

In the following, we will use  $\INDICATRICE{Q}{\EVENTO}$ to represent
the indicator function {of  node set  $Q$} taking value $1$ if $Q\cap\EVENTO\neq\emptyset$, and value $0$ otherwise.
Accordingly, we have that the probability $\PNODE{s}$, 
that node $s\in\SET[T]$ issues a request, is 
$\PNODE{s}=\sum_{\EVENTO\subseteq\SET[T]}\INDICATRICE{\{s\}}{\EVENTO}\mathbb{P}(\EVENTO)$ whereas  
the probability that at least one node in a set of nodes $S\subseteq\SET[T]$ issues a request is $\PSETS{S}=\sum_{\EVENTO\subseteq\SET[T]}\INDICATRICE{S}{\EVENTO}\mathbb{P}(\EVENTO)$ or equivalently $\PSETS{S}=1 - \PNOSETS{S}$. Naturally, if $S=\emptyset$ then $\PSETS{S}=0$.
Table \ref{tab:notation} summarizes the main notation used to define the problem. 

\begin{table}[!htb]
	\centering
	%\begin{tabular}{|c|c|}
	\begin{tabular}{|p{4cm}|p{10cm}|}
		\hline 
		\textbf{Notation}  &  \textbf{Description} \\
		\hline \hline 
		\TREET\ & main tree on which the problem is defined\\ \hline
		\SET[T] & node set of the main tree \TREET\\ \hline
		\ROOT[T] & root of the main tree \TREET\\ \hline
\TREEA\ (\SUBTREE{A}) &  generic sub-tree of \TREET\ ($i$-th sub-tree of \TREEA)\\ \hline
\SET\ (\SUBSET{A})&  set of nodes of the generic tree  \TREEA\ (of $i$-th sub-tree of \TREEA)\\ \hline
%\SUBSET{A})&  set of nodes of the generic tree  \TREEA\ (of $i$-th sub-tree of \TREEA)\\ \hline
$\ROOT, \PROOTA,
\PRIZE{\ROOT[A]},
\DISTA$ & root of \TREEA, probability, prize and distance from the root of \TREET\\ \hline
\ARC[i]  % \ARC[i]{A} 
& weight of the edge connecting  root \ROOT\ of    \TREEA\ with the root of $i$-th sub-tree \SUBTREE{A}\\ \hline
 $\TREEINDEX=\{0,1,\ldots,m\}$ & index set of the sub-trees of \TREEA\\ \hline
$\SETAINDEXED,$ $I\subseteq\TREEINDEX$ & set of nodes belonging to
a subset $I$ of sub-trees of \TREEA\\  \hline
 $\PSETS{S} = 1 - \PNOSETS{S}$ & probability that at least one node in the set  $S$ offers a non-null prize\\	  \hline
	\end{tabular}
	\caption{Main notation}\label{tab:notation}
\end{table}

\subsection{Problem definition}

Let $S\subseteq\SETT$ be a set of nodes on which an agreement has been a priori established
%, so that the manager is committed to serve all nodes in $S$ issuing a request for the corresponding prizes. In all cases, nodes not in $S$ will not be served nor will they produce a reward for the manager.
%
and let $\EVENTO\subseteq\SETT$ be the set of nodes issuing a request of service ({\it nature outcome}).
The nodes that have to be visited are then given by $S\cap\EVENTO$. %\subseteq\SETT$.

\smallskip
According to the a priori selection of $S$, let us define three random variables  all depending on the outcome \EVENTO: the \textit{revenue}, the \textit{cost}, and the \textit{profit} defined as difference between revenue and cost.
%, all depending on the outcome \EVENTO.
%
The distributions, and in particular the expected value of these random variables, depend on the node set $S$ selected a priori and their analytical expressions follow.
%
%\red{The PPTP-T looks for a selection of $S$ so  that tour that maximize visiting objective is the maximization of the expected value of the profit.}
%\note{????}
%
%The analytical expressions of these random variables follow.

\subsubsection{Revenue}
The revenue produced by a node set $S$ is defined as %can be expressed both explicitly and recursively as indicated below
\begin{align*}
	\RRICAVO{S}      &= \sum_{s\in S}\PRIZE{s}\INDICATRICE{\{s\}}{\EVENTO},
\end{align*}
and if $S\subseteq\SET$ for a sub-tree \TREEA, then the revenue can be recursively computed as
\begin{align*}
	\RRICAVO{S} = \RRICAVO{S\cap\SET}     & =\SUMTREE\RRICAVO{S\cap\SUBSET{A}}.
\end{align*}%
The expected value of the revenue produced by node set $S$ is thus given by % can be expressed both explicitly and recursively as indicated below
\begin{align*}
	\MRICAVO{S} \equiv \VALOREATTESO{\RRICAVO{S}} & =  
	    \sum_{\EVENTO\subseteq\SETT}\left(\sum_{s\in S}\PRIZE{s}\INDICATRICE{\{s\}}{\EVENTO}\right)\mathbb{P}(\EVENTO)
	    =\sum_{s\in S}\PRIZE{s}\left(\sum_{\EVENTO\subseteq\SETT}\INDICATRICE{\{s\}}{\EVENTO}\mathbb{P}(\EVENTO)\right)
	    =\sum_{s\in S}\PRIZE{s}\PNODE{s},
\end{align*}
and by
\begin{align*}
	\MRICAVO{S} = \MRICAVO{S\cap\SET} & = \SUMTREE\MRICAVO{S\cap\SUBSET{A}}
\end{align*}
when $S\subseteq\SET$ for a sub-tree \TREEA.

\subsubsection{Discounted cost}

To describe the cost random variable, we introduce a {\it bonus} $x$ representing a discount applied to the cost of the common path starting from the root of the main tree \TREET\ and shared by all paths reaching nodes in the selected set  $S$.
This bonus, not strictly relevant to the problem definition, is used as a tool to define the optimal solutions and to prove problem  complexity, as showed  in the following.

Thus, let $S$ be an a priori node selection and \TREEA\ any sub-tree of \TREET\ such that $S\subseteq\SET[A]$.
For all $x\in[0,\DISTA]$, the random discounted cost implied by $S$ with bonus $x$ is 
\begin{align}
	\RCOSTO{S}{x}    &= \left(\DISTA-x+\RCOSTO{S}{\DISTA}\right)\INDICATRICE{S}{\EVENTO} \label{RCOSTO_def}
\end{align}

The random variable \RCOSTO{S}{x} indicates  the minimum traveling cost of a tour visiting  nodes in $S\cap\EVENTO$ from root \ROOT[T] when we assume to have a bonus equal to $x$. This cost function  
is null if $\INDICATRICE{S}{\EVENTO}=0$ (i.e. $S\cap\EVENTO=\emptyset$), otherwise it is computed as the cost to reach the root \ROOT\ of \TREEA\ (common to all the nodes in $S\cap\EVENTO$) net of the bonus $x$ plus the cost  
to visit the nodes in $S\cap\EVENTO$ computed starting from the root of  \TREEA.
%\Renata
%
%Si osservi la "sinergia" fra i nodi di $S$ sulla parte comune del percorso a partire da $x$ il cui costo è conteggiato una sola volta e quindi condiviso.
%
Regarding the term $\RCOSTO{S}{\DISTA}$, it
can be exploded into the sum of  costs incurred in the different branches of tree \TREEA: 
\begin{equation}
	\RCOSTO{S}{\DISTA}=\SUMTREE\RCOSTO{S\cap\SUBSET{A}}{\DISTA}. \label{eq:costo_con_sviluppo_sottoalberi} %{cost_def}
\end{equation}
Indeed, nodes in $S\cap\EVENTO$ jointly contribute to cover 
the common cost $\DISTA -x$ to reach the root of  \TREEA, whereas they no longer have anything in common starting from the root  \ROOT\ and therefore no common cost to share.
If we use $\ARC[i]=\DIST{\SUBROOT{A}}-\DISTA$ and apply definition \eqref{RCOSTO_def}
recursively on \eqref{eq:costo_con_sviluppo_sottoalberi} % $\RCOSTO{S\cap\SUBSET{A}}{\DISTA}$ 
we get
$$
%\RCOSTO{S}{\DISTA}=\SUMTREE\left(\ARC{A}+\RCOSTO{S\cap\SUBSET{A}}{\DIST{\SUBROOT{A}}}\right)\INDICATRICE{S\cap\SUBSET{A}}{\EVENTO}.
\RCOSTO{S}{\DISTA}=\SUMTREE\left(\ARC+\RCOSTO{S\cap\SUBSET{A}}{\DIST{\SUBROOT{A}}}\right)\INDICATRICE{S\cap\SUBSET{A}}{\EVENTO}
$$
from which we can observe that 
if $S\cap\EVENTO=\emptyset$, it necessarily holds that $\INDICATRICE{S\cap\SUBSET{A}}{\EVENTO}=0$ for each $i\in\TREEINDEX$
and thus $\RCOSTO{S}{\DISTA}$ is null without requiring to be multiplied by  $\INDICATRICE{S}{\EVENTO}=0$. 
As a consequence, we can rewrite 
expression \eqref{RCOSTO_def} as follows:

\begin{align}
	\RCOSTO{S}{x}=(\DISTA-x)\INDICATRICE{S}{\EVENTO}+\RCOSTO{S}{\DISTA}.\label{eq:costo_con_sviluppo_bonus}
\end{align}

It is worth noticing that, as stated in the following proposition, 
computing $\RCOSTO{S}{x}$ returns always the same value for any tree \TREEA\ such that $S\subseteq\SET$ and  $x\in[0,\DISTA]$:

\begin{proposition}\label{P:costo_atteso_ben_definito}
	Let \TREEA\ and  \TREEB\ be two trees such that  $S\subseteq\SET[A]$ and  $S\subseteq\SET[B]$ and $x\leq\min(\DISTA,\DISTB)$. Then, for each outcome  \EVENTO, it holds that
	\begin{align*}
		(\DISTA-x)\INDICATRICE{S}{\EVENTO}+\RCOSTO{S}{\DISTA}=(\DISTB-x)\INDICATRICE{S}{\EVENTO}+\RCOSTO{S}{\DISTB}.
	\end{align*}
\end{proposition}
\begin{proof}[Proof of Proposition \ref{P:costo_atteso_ben_definito}]
	If  $S=\emptyset$ the equality trivially holds with $$(\DISTA-x)\INDICATRICE{S}{\EVENTO}+\RCOSTO{S}{\DISTA}=(\DISTB-x)\INDICATRICE{S}{\EVENTO}+\RCOSTO{S}{\DISTB}=0.$$
	Otherwise, one tree contains the other as a sub-tree. Without any loss of generality, let us assume that \TREEB\ is contained in \TREEA\ and thus $\DISTA\leq\DISTB$. 
	%
	%
	%
	%\note{Dovrebbe essere chiaro che per noi un insieme di sottoalberi non è un sottoalbero.}
	%
	%
	%
	Notice that as $S\subseteq\SET[B]$ and \TREEB\ is a subtree of \TREE, $S$ can be contained in only one sub-tree \SUBTREE[h]{A} of \TREEA\ and thus  $S\cap\SUBSET[h]{A}=S$ and $S\cap\SUBSET[i]{A}=\emptyset$ for all the remaining sub-trees $i \neq h, i \in \TREEINDEX$.

	Starting from  \eqref{eq:costo_con_sviluppo_bonus} and \eqref{eq:costo_con_sviluppo_sottoalberi} and further developing them  we get:
	\begin{align*}
		\RCOSTO{S}{x} 
		& =(\DISTA-x)\INDICATRICE{S}{\EVENTO}+\RCOSTO{S}{\DISTA}\\
		& =(\DISTA-x)\INDICATRICE{S}{\EVENTO}+\SUMTREE\RCOSTO{S\cap\SUBSET{A}}{\DISTA}\\
		& =(\DISTA-x)\INDICATRICE{S}{\EVENTO}+\RCOSTO{S\cap\SUBSET[h]{A}}{\DISTA}\\
		& =(\DISTA-x)\INDICATRICE{S}{\EVENTO}+(\DIST{\SUBROOT[h]{A}}-\DISTA)\INDICATRICE{S}{\EVENTO}+\RCOSTO{S}{\DIST{\SUBROOT[h]{A}}}\\
		& =(\DIST{\SUBROOT[h]{A}}-x)\INDICATRICE{S}{\EVENTO}+\RCOSTO{S}{\DIST{\SUBROOT[h]{A}}}\\
		& = \RCOSTO{S}{x},
	\end{align*}%
	{observe that the first and last occurrences of $\RCOSTO{S}{x}$ correspond to computing, by formula \eqref{eq:costo_con_sviluppo_bonus}, the realized cost from \TREE\ and its sub-tree \SUBTREE[h]{A}, respectively.}
	
	If $\TREEB=\SUBTREE[h]{A}$ we have done, otherwise we should iterate down in the sub-tree \SUBTREE[h]{A} up to \TREEB\ keeping constant the value of \RCOSTO{S}{x}.
\end{proof}

\medskip

We can conclude that the random variable \RCOSTO{S}{x} is well 
defined for each value of $x$ provided that 
there exists a tree \TREEA\ such that $S\subseteq\SET$ and $x\leq\DISTA$.
The expected value of the cost function can thus be computed as follows:

\begin{align}
	\MCOSTOB{S}{x} \equiv \VALOREATTESO{\RCOSTO{S}{x}} & =(\DISTA-x)\PSETS{S}+\MCOSTOB{S}{\DISTA}\label{eq:costo_atteso_con_sviluppo_bonus}\\
	& = (\DISTA-x)\PSETS{S}+\SUMTREE\MCOSTOB{S\cap\SUBSET{A}}{\DISTA}.\label{eq:costo_atteso_con_sviluppo_sottoalberi}
\end{align}

Expression \eqref{eq:costo_atteso_con_sviluppo_bonus} immediately follows from:  
\begin{align*}
	\VALOREATTESO{\RCOSTO{S}{x}}
	& =\sum_{\EVENTO\subseteq\SET[T]}\left[(\DISTA-x)\INDICATRICE{S}{\EVENTO}+\RCOSTO{S}{\DISTA}\right]\mathbb{P}(\EVENTO)\\
	& =(\DISTA-x)\sum_{\EVENTO\subseteq\SET[T]}\INDICATRICE{S}{\EVENTO}\mathbb{P}(\EVENTO)+\sum_{\EVENTO\subseteq\SET[T]}\RCOSTO{S}{\DISTA}\mathbb{P}(\EVENTO)\\
	& =(\DISTA-x)\PSETS{S}+\MCOSTOB{S}{\DISTA},
\end{align*}

whereas \eqref{eq:costo_atteso_con_sviluppo_sottoalberi} can be proved as follows:

\begin{align*}
	\VALOREATTESO{\RCOSTO{S}{\DISTA}}
	& =\sum_{\EVENTO\subseteq\SET[T]}\left[\SUMTREE\RCOSTO{S\cap\SUBSET{A}}{\DISTA}\right]\mathbb{P}(\EVENTO)\\
	& =\SUMTREE\left[\sum_{\EVENTO\subseteq\SET[T]}\RCOSTO{S\cap\SUBSET{A}}{\DISTA}\mathbb{P}(\EVENTO)\right]\\
	& =\SUMTREE\MCOSTOB{S\cap\SUBSET{A}}{\DISTA}.
\end{align*}

\subsubsection{Profit}
Finally, 
for any node set $S$ and $x\in[0,\DISTA]$ for some sub-tree \TREEA\ such that $S\subseteq\SET[A]$,
the random variable profit implied by $S$ is 
%is defined as
\begin{align*}
	\RGUADAGNO{S}{x} &= \RRICAVO{S}-\RCOSTO{S}{x} 
\end{align*}
to represent the profit gained to serve nodes $S\cap\EVENTO$ with a cost discount $x$. 
%Following what we proved above 
Hence, the expected value of \RGUADAGNO{S}{x} can be computed as follows:
\begin{align}
	\nonumber \MGUADAGNOB{S}{x} \equiv \VALOREATTESO{\RGUADAGNO{S}{x}} & = \MRICAVO{S} - \MCOSTOB{S}{x}\\
	& = \MGUADAGNOB{S}{\DISTA}-(\DISTA-x)\PSETS{S}\label{eq:guadagno_atteso_con_sviluppo_bonus}\\
	& = \SUMTREE\MGUADAGNOB{S\cap\SUBSET{A}}{\DISTA}-(\DISTA-x)\PSETS{S}.\label{eq:guadagno_atteso_con_sviluppo_sottoalberi}
\end{align}

PPTP-T looks for a set $S$ that maximizes  the expected value of the profit random variable \RGUADAGNO{S}{x} when the bonus $x$ is null:
%. It can be formalized as follows: 

\begin{equation}
	z^*=\max_{S\subseteq\SET[T]}\VALOREATTESO{\RGUADAGNO{S}{0}}.\label{DEF:PROBLEMA_STOCASTICO_BASE_BONUS}
\end{equation}

Solving problem \eqref{DEF:PROBLEMA_STOCASTICO_BASE_BONUS} may appear a difficult task as the number of potential options is exponentially large in the size of the  node set \SET[T].
To solve it, we 
%\remove{are going to introduce a 
%\red{dynamic programming approach} that requires, for each subset $I$ of sub-trees of \TREEA\ ($I\subseteq\TREEINDEX$), to}
study the  properties and generation of a function \FCSETINDEXA{I} defined as follows:
\begin{equation}\label{eq:funzione_caratteristica_sottoalberi}
	\FCSETINDEXA{I}=\max_{S\subseteq\SETAINDEXED}\MGUADAGNOB{S}{x}.
\end{equation}

We call this function the \textit{\SUBTREECHARFUNCTION} with respect to the group of sub-trees  $I$, whereas a solution of the corresponding optimization problem is called \textit{\OPTIMALSET}.
Analogously, we indicate as $\FCSETINDEXA{\TREEINDEX}$ 
the  \textit{\TREECHARFUNCTION} of  \TREEA\ which, for sake of simplicity, will be indicated as 
\begin{equation}\label{eq:funzione_caratteristica}
	\FCA
\end{equation}
It follows that:
$$
z^*=f^{(\TREET)}(0).
$$

In the following sections, we show that problem \eqref{DEF:PROBLEMA_STOCASTICO_BASE_BONUS}
can be optimally solved in time  $O(|\SET[T]|^2)$.

\section{Properties}\label{SEC:properties}
In this section, we discuss the properties of the \SUBTREECHARFUNCTION\ for a generic tree \TREE\ and any group of sub-trees $I\subseteq\TREEINDEX$.

First of all, we show that \FCSETINDEXA{I} is a monotone non-decreasing piece-wise linear function with a number of linear traits not larger than $|\SETAINDEXED|+1$ over the domain $[0,\DISTA]$ of the bonus $x$.

In Section \ref{SEC:CHARACTERISTIC_FUNCTION} we show that each linear segment
is characterized by a unique optimal set, each corner point
%(the points connecting each linear segment to the subsequent one)
is characterized by a unique minimal optimal set (left segment) and a unique maximal optimal set (right segment),
%. In particular, the minimal optimal set is the unique optimal set to the left of the corner point and the maximal optimal set is the unique optimal set to its right.
%
and that the sequence of \OPTIMALSETS\ defines a sort of {\it matryoshka} which grows while the bonus moves from left to right in the domain of \FCSETINDEXA{I}.

On this basis, in Section \ref{SEC:Computing_FCA} we finally show the properties of \SUBTREECHARFUNCTIONS\ \FCSETINDEXA{I} that can be exploited to efficiently compute the \TREECHARFUNCTION\ \FCA.

\bigskip
Let us start by observing that, given a tree \TREEA\ and a group of sub-trees $I\subseteq\TREEINDEX$, the expected profit of a set of nodes $S\subseteq\SETAINDEXED[I]\subseteq\SET$ can be expressed, for all $x\in[0,\DISTA]$, by means of formula \eqref{eq:guadagno_atteso_con_sviluppo_bonus} as 
\begin{align}
	\MGUADAGNOB{S}{x} &= \left(\MGUADAGNOB{S}{\DISTA}-\PSETS{S}\cdot\DISTA\right)+\PSETS{S}\cdot x\label{eq:profitto_atteso_Lineare_rispetto_al_Bonus}
\end{align}
to emphasize its dependency on the bonus $x$. 
We call \eqref{eq:profitto_atteso_Lineare_rispetto_al_Bonus} the \textit{\PROFITLINE} of node set $S$.
Note that the \PROFITLINE\ of a node set  $S$ can be graphically represented as a line with non negative slope equal to $\PSETS{S}$, that is the probability that at least one node in $S$ issues a request.
In particular, the slope is  positive for each set  $S\neq\emptyset$, and null for
$S=\emptyset$ only; in such a case,  $\MGUADAGNOB{\emptyset}{x}=0$ for all  $x\in[0,\DISTA]$.

\smallskip
Thus, from the definition of function \FCSETINDEXA{I} as the superior envelope of a finite number of linear segments with non negative slopes, we can claim the following fact:

\begin{fact}
	For any tree \TREEA\ and group of sub-trees $I\subseteq\TREEINDEX$, the function \FCSETINDEXA{I} is well defined for $x\in[0,\DISTA]$, and is a continuous, convex, non-decreasing, piece-wise linear function. Each linear piece is the \PROFITLINE\ of an \OPTIMALSET\ within the corresponding range. 
\end{fact}

\subsection{{Properties of the} \SUBTREECHARFUNCTION}\label{SEC:CHARACTERISTIC_FUNCTION}
\paragraph{Maximal and minimal \OPTIMALSETS.}%\ for the \SUBTREECHARFUNCTION\ \FCSETINDEXA{I}.}
Given a tree \TREEA\ and a group  $I\subseteq\TREEINDEX$ of sub-trees, 
 for each value of the bonus $x\in[0,\DISTA]$,  there might be several \OPTIMALSETS\ {defining the \SUBTREECHARFUNCTION\ \FCSETINDEXA{I}}. 
In Proposition  
\ref{P:prop_Unicita_Ottimo_Massimale_Minimale}, {we show}
that 
 there is only one \MINOPTIMALSET\ and only one \MAXOPTIMALSET. %, possibly coincident.
Before discussing Proposition \ref{P:prop_Unicita_Ottimo_Massimale_Minimale}, we need the following technical result whose proof is provided in Appendix.

\begin{proposition}\label{P:prop_operazioni_insiemi}
	Let $S_1$ and $S_2$ be two node sets and \TREEA\ a tree containing both $S_1$ and $S_2$ (i.e. $S_1,S_2\subseteq\SET$). Then for any $x\in[0,\DISTA]$ we have 
	
	\begin{align}
		\MGUADAGNOB{S_1}{x}+\MGUADAGNOB{S_2}{x} & \leq \MGUADAGNOB{S_1\cup S_2}{x}+\MGUADAGNOB{S_1\cap S_2}{x}.\label{eq:prop_Profitto_su_operazioni_insiemi}
	\end{align}
\end{proposition}

%\begin{remark} Propositions \ref{P:prop_subset_cost_inequality} and \ref{P:prop_operazioni_insiemi} tell us that if a node $v$ has a negative prize $\PRIZE{v}<0$, then it cannot be part of an \OPTIMALSET. In fact given a node set $S$ not containing $v$, we get from \eqref{eq:prop_Ricavo_su_operazioni_insiemi} $\MRICAVO{S\cup\{v\}}=\MRICAVO{S}+\MRICAVO{\{v\}}< \MRICAVO{S}$; while from Proposition we have \ref{P:prop_subset_cost_inequality} $\MCOSTOB{S\cup\{v\}}{x}\geq\MCOSTOB{S}{x}$. Inequality $\MGUADAGNOB{S}{x}>\MGUADAGNOB{S\cup\{v\}}{x}$ follows.\end{remark}

\begin{proposition}\label{P:prop_Unicita_Ottimo_Massimale_Minimale}
	Given a tree \TREEA, a group of sub-trees $I\subseteq\TREEINDEX$ and a bonus $x\in[0,\DISTA]$, there exists a unique \MINOPTIMALSET\ and a unique \MAXOPTIMALSET\ corresponding to \FCSETINDEXA{I}. 
	%Moreover, any \OPTIMALSET\ is a subset of the maximal one and a superset of the minimal one.
\end{proposition}

\begin{proof}[Proof of Proposition \ref{P:prop_Unicita_Ottimo_Massimale_Minimale}]
    First of all, we observe that given two distinct \OPTIMALSETS\ $S_1$ and $S_2$ for the same bonus $x$ (i.e. $\MGUADAGNOB{S_1}{x}=\MGUADAGNOB{S_2}{x}$=\FCSETINDEXA{I}), then also $S_1\cup S_2$ and $S_1\cap S_2$ are \OPTIMALSETS.
    
    Indeed, we have 
        $\MGUADAGNOB{S_1}{x}=\MGUADAGNOB{S_2}{x}\leq\max(\MGUADAGNOB{S_1\cup S_2}{x},\MGUADAGNOB{S_1\cap S_2}{x})$,
    otherwise we get
	    $\MGUADAGNOB{S_1}{x}+\MGUADAGNOB{S_2}{x}>\MGUADAGNOB{S_1\cup S_2}{x}+\MGUADAGNOB{S_1\cap S_2}{x}$
    in contrast with \eqref{eq:prop_Profitto_su_operazioni_insiemi}.
    For the optimality of $S_1$ and $S_2$ we conclude that
	%$\MGUADAGNOB{S_1}{x}=\MGUADAGNOB{S_2}{x}=\max(\MGUADAGNOB{S_1\cup S_2}{x},\MGUADAGNOB{S_1\cap S_2}{x})$, which proves that 
	at least one between  $S_1\cap S_2$ and $S_1\cup S_2$ is optimal. 
	Actually, they are both optimal because from inequality \eqref{eq:prop_Profitto_su_operazioni_insiemi} we know that
	\[
	\MGUADAGNOB{S_1}{x}+\MGUADAGNOB{S_2}{x}\leq\MGUADAGNOB{S_1\cup S_2}{x}+\MGUADAGNOB{S_1\cap S_2}{x}),
	\]
	and subtracting on both side the optimal value (assume w.l.o.g. $\MGUADAGNOB{S_1}{x}=\MGUADAGNOB{S_1\cup S_2}{x}$) we remain with 
	\[
	\MGUADAGNOB{S_2}{x}\leq\MGUADAGNOB{S_1\cap S_2}{x},
	\]
	which, from the optimality of $S_2$, also proves the optimality of $S_1\cap S_2$.

    \medskip
    Now we proceed by contradiction,
	let $S_1$ and $S_2$ be two distinct \MAXOPTIMALSETS\ (i.e. $S_1\not\subseteq S_2$ and $S_2\not\subseteq S_1$), we know that their proper superset $S_1\cup S_2$ and their proper subset $S_1\cap S_2$ are optimal, thus neither $S_1$ nor $S_2$ can be a \MAXOPTIMALSET\ or a \MINOPTIMALSET.
\end{proof}

\paragraph{Matryoshka property.}% of the \SUBTREECHARFUNCTION\ \FCSETINDEXA{I}.}%
In Propositions \ref{P:FC_punto_angoloso}, we show that two distinct \OPTIMALSETS\ for \FCSETINDEXA{I} may exist only if $x$ is a corner point.
Moreover the unique \MINOPTIMALSET\ (guaranteed by Proposition \ref{P:prop_Unicita_Ottimo_Massimale_Minimale}) is the unique \OPTIMALSET\ for the linear trait on the left of the corner point, and, analogously, the unique \MAXOPTIMALSET\ is the unique \OPTIMALSET\ on  the linear trait on the right of the corner point. % (Proposition \ref{P:FC_punto_angoloso}). 
Finally, in Proposition \ref{P:FC_numero_punti_angolosi}, we show that the \OPTIMALSETS\ in the domain $[0,\DISTA]$ form a matryoshka implying that function \FCSETINDEXA{I} has at most $|\SETAINDEXED|$ corner points and $|\SETAINDEXED|+1$ linear traits (possibly including the empty set). % (Proposition \ref{P:FC_numero_punti_angolosi})

\begin{proposition}\label{P:FC_punto_angoloso}
	Let us consider a tree \TREEA\ and a group of sub-trees $I\subseteq\TREEINDEX$. 
	If, for some $\bar{x}\in(0,\DISTA)$, there exist two distinct \OPTIMALSETS, then there is only one \MAXOPTIMALSET\ which is optimal in $[\bar{x},\bar{x}+\varepsilon)$ and only one \MINOPTIMALSET\ which is optimal in $(\bar{x}-\varepsilon,\bar{x}]$ for some $\varepsilon>0$. In particular, $\bar{x}$ is a corner point of \FCSETINDEXA{I}.
\end{proposition}

\begin{proof}[Proof of Proposition \ref{P:FC_punto_angoloso}]
	From Proposition \ref{P:prop_Unicita_Ottimo_Massimale_Minimale} we know that, for each given $x\in[0,\DISTA]$, a unique minimal and a unique \MAXOPTIMALSET\ exist, the first subset of the second one.
	If, for a given $\bar{x}\in(0,\DISTA)$ we have two distinct \OPTIMALSETS, then the minimal and maximal ones are distinct, and the \MINOPTIMALSET\ is a proper subset of the \MAXOPTIMALSET. Let us call them $Y$ and $Z$, respectively. 
	Clearly, any other \OPTIMALSET\ $W$ is a proper superset of $Y$ and a proper subset of $Z$.
	It is evident from formula \eqref{eq:profitto_atteso_Lineare_rispetto_al_Bonus} that the slopes of their \PROFITLINES\ are $\PSETS{Y}<\PSETS{W}<\PSETS{Z}$ (larger node sets manifest with higher probability). 
	Thus, from $\MGUADAGNOB{Y}{\bar{x}}=\MGUADAGNOB{W}{\bar{x}}=\MGUADAGNOB{G}{\bar{x}}$, we immediately derive $\MGUADAGNOB{Y}{x}>\MGUADAGNOB{W}{x}>\MGUADAGNOB{Z}{x}$ for $x<\bar{x}$ and vice-versa $\MGUADAGNOB{Y}{x}<\MGUADAGNOB{W}{x}<\MGUADAGNOB{Z}{x}$ for $x>\bar{x}$.
	For any other non \OPTIMALSET\ $S$ in $\bar{x}$, we know that $\MGUADAGNOB{S}{\bar{x}}<\MGUADAGNOB{Y}{\bar{x}}$, and the inequality holds in a neighborhood of $\bar{x}$ for the continuity of the  \PROFITLINES.
	Finally, if for a fixed $\varepsilon>0$ there were some distinct \OPTIMALSETS\ in interval $(\bar{x}-\varepsilon,\bar{x}+\varepsilon)$ other than $Y$ and $Z$, then we can repeatedly halve the value of $\varepsilon$; the process must come to an end as we have only a finite number of potential \OPTIMALSETS.
	Point $\bar{x}$ is thus a corner point since function \FCSETINDEXA{I} takes different slopes around $\bar{x}$.
\end{proof}

\begin{proposition}\label{P:FC_numero_punti_angolosi}
	Given a tree \TREEA\ and a group of sub-trees $I\subseteq\TREEINDEX$, 
	the domain $[0,\DISTA]$ of function \FCSETINDEXA{I} contains at most $|\SETAINDEXED|$ corner points and $|\SETAINDEXED|+1$ linear pieces. 
	Furthermore, the \OPTIMALSETS\ corresponding to the sequence of linear traits form a chain with respect to inclusion (matryoshka).
\end{proposition}

\begin{proof}[Proof of Proposition \ref{P:FC_numero_punti_angolosi}]
	According to Proposition \ref{P:FC_punto_angoloso}, at each corner point the \OPTIMALSET\ gains additional nodes, and since we have $|\SETAINDEXED|$ nodes, we can have $|\SETAINDEXED|$ corner points at most. 
	Starting from the empty set the number of \OPTIMALSETS\ cannot be greater than $|\SETAINDEXED|+1$. 
	Finally, moving from left to right in $[0,\DISTA]$, at each corner point we add one or more new nodes to the previous \OPTIMALSET\ so that a the new \OPTIMALSET\ is a superset of the previous one.
\end{proof}

\subsection{Computing \FCA}\label{SEC:Computing_FCA}
%
%In this Section we propose a data structure \DESCRIZIONEAI\ to describe \FCSETINDEXA{I} as a list of at most $|\SETAINDEXED|+1$ records, and then illustrate the properties that allow us to efficiently compute it for any tree \TREE\ and group $I$ of sub-trees.
%
%This  goal is achieved by computing \FCSETINDEXA{I} for  ever larger sub-tree groups until we have $I=\TREEINDEX$. 

In this section, we propose a data structure \DESCRIZIONEAI\ to describe \FCSETINDEXA{I} as a list of at most $|\SETAINDEXED|+1$ records, and then illustrate the properties that allow us to efficiently compute \FCA\ for any tree \TREE.
This  goal is achieved by a bottom-up approach starting from the \TREECHARFUNCTION\ \FCA\ of trees without descendants (leaf nodes of \TREET) and by aggregating (or merging) ever larger groups of sub-trees until we have \FCSETINDEXA{I} for $I=\TREEINDEX$. 
Discussion will go through the following steps: %Main steps of the discussion will be the following
\begin{itemize}
    \item building the description of \FCA\ starting from leaf sub-trees;% and proceed by aggregation of all sub-trees in the same tree;
    \item the descriptions \DESCRIZIONEAI[I'] and \DESCRIZIONEAI[I''] of the \SUBTREECHARFUNCTION\ of two disjoint groups of sub-trees $I'$ and $I''$ can be "merged" into the description \DESCRIZIONEAI[I'\cup I''] of the union group $I'\cup I''$ by exploring the Cartesian product of the two descriptions instead of the power set of \SETAINDEXED[I'\cup I''];
    
%    \item \red{ ALTERNATIVA EQUIVALENTE ALLA PRECEDENTE: given two disjoint groups of sub-trees $I',I''\subseteq\TREEINDEX$, the description \DESCRIZIONEAI[I'\cup I''] of the union group $I'\cup I''$ can be obtained by exploring the cartesian product of the two descriptions \DESCRIZIONEAI[I'] and \DESCRIZIONEAI[I''] instead of the power set of \SETAINDEXED[I'\cup I''];}

    \item the computation of the description \DESCRIZIONEAI[I'\cup I''] of \FCSETINDEXA{I'\cup I''} can be even more efficient with an appropriate exploration strategy of the descriptions of \FCSETINDEXA{I'} and \FCSETINDEXA{I''}.
\end{itemize}

\subsubsection{Characteristic function representation}\label{SEC:MATRIOSKA}

Given a tree \TREEA\ and a group of its sub-trees  $I\subseteq\TREEINDEX$, we refer to  
$$
\RDESCRIZIONEAI{i} =\langle \FIELDSET, \FIELDXMIN, \FIELDXMAX, \FIELDPENDENZA, \FIELDQZERO \rangle
$$
as the tuple of the fields fully describing  the $i$-th linear piece of the  \SUBTREECHARFUNCTION\ \FCSETINDEXA{I} on the domain $[0,\DISTA]$.

In particular,
\FIELDSET\ indicates the  \MAXOPTIMALSET\ on the interval $[\FIELDXMIN,\FIELDXMAX)$, whereas
\FIELDPENDENZA\ and \FIELDQZERO\ represents the slope of the \SETCHARFUNCTION\ associated with  \FIELDSET\ (i.e. \PSETS{S}) and the value of the \PROFITLINE\ when $x=\DISTA$ (i.e. \MGUADAGNOB{S}{\DISTA}).
The complete description of the linear pieces of \FCSETINDEXA{I} is thus given by a vector of tuples 
$$
\DESCRIZIONEAI = [\RDESCRIZIONEAI{i}]_{i=0}^{\CARDA}
$$
where the  $\CARDA+1\leq|\SETAINDEXED|+1$ records are sorted in such a way that 
 $\RDAXMINI{i} < \RDAXMINI{i+1}$ for each  $i=0,1,\ldots,\CARDA-1$. 
Then, it follows that
\begin{align*}
	\RDAXMAXI{i} & = \RDAXMINI{i+1}\\
	\RDASETI{i} & \subset\RDASETI{i+1}\\
	\RDAPENDENZAI{i} & < \RDAPENDENZAI{i+1}
\end{align*}
and, in particular,  $\RDAXMINI{0}=0$, $\RDAXMAXI{\CARDA}=\DISTA$. 
Observe that, accordingly to the previous discussion, \RDASETI{i} is 
the only \MAXOPTIMALSET\ for $x=\RDAXMINI{i}$,
the only \OPTIMALSET\ for $x\in (\RDAXMINI{i},\RDAXMAXI{i})$,
the only \MINOPTIMALSET\ for $x=\RDAXMAXI{i}$.
Also note that, 
while $\RDAXMINI{i}<\RDAXMAXI{i}$ for all $i<\CARDA$,
it could be given the case that range $[\RDAXMINI{\CARDA},\RDAXMAXI{\CARDA}]$ reduces to a single point if a new node enters the \MAXOPTIMALSET\ exactly when $x=\DISTA$. 
Finally, if $\DISTA=0$ we have $\CARDA=0$ and only one record \RDESCRIZIONEAI{0} to describe \FCA\ in the domain $[0,0]$.
For presentation convenience, %if not ambiguous in the text, 
we %will
indicate with
% \MATSETI{i} the $i$-th \OPTIMALSET\ \RDASETI{i} implied by sub-trees group $I\subseteq\TREEINDEX$, and with 
%\begin{align*}
%	\MATRIOSKAAI \equiv [\MATSETI{0} \subset \MATSETI{1} \subset \MATSETI{2} \subset\ldots \subset \MATSETI{\CARDA}]
%\end{align*}
\begin{align*}
	\MATRIOSKAAI \equiv [\RDASETI{0} \subset \RDASETI{1} \subset \RDASETI{2} \subset\ldots \subset \RDASETI{\CARDA}]
\end{align*}
the nested %matryoshka 
sequence of the \OPTIMALSETS\ (matryoshka).%\ of the group $I$ of sub-trees of \TREEA. 

Note that all the information in \DESCRIZIONEAI\ is directly implied by the matryoshka %\remove{sequence} 
\MATRIOSKAAI\ from which all other information can be computed. 
This is the reason why we focus only on the calculation of the matryoshka {\MATRIOSKAAI}.
Nevertheless, for computational efficiency reasons, it is convenient to explicitly store all the elements of the tuple $\RDESCRIZIONEAI{i}$ how illustrated in Section \ref{SEC:algorithm} where the algorithm implementation is commented.

For sake of simplicity, we remove index $I$ from notation when the group of sub-trees is complete (i.e.\ $I=\TREEINDEX$) and use \DESCRIZIONEA\ and \MATRIOSKAA\ instead of  \DESCRIZIONEAI[\TREEINDEX] and \MATRIOSKAAI[\TREEINDEX], in the same way as \FCA\ is an alias for \FCSETINDEXA{\TREEINDEX}.

\subsubsection{The basic case (\TREE\ is a leaf)}\label{SEC:MATRIOSKA2}

To compute  \FCA\ for a tree \TREEA\ without descendants, the decision is only whether or not to select the root of the tree. 
The choice is determined by the following expression:
\begin{align*}
	\FCA & =\max(\MGUADAGNOB{\emptyset}{x},\MGUADAGNOB{\{\ROOT\}}{x})\\
	& =\max(0,(\PRIZE{\ROOT}-(\DISTA-x))\PROOTA)\\
	& =\left\{\begin{array}{ll}(x-(\DISTA-\PRIZE{\ROOT}))\PROOTA & \text{ if }x\geq\DISTA-\PRIZE{\ROOT}\\ 0 & \text{ otherwise.}\end{array}\right.
\end{align*}
Thus, the description of  %\FCA\ 
\DESCRIZIONEA\ is given by 
\begin{align*}
	\DESCRIZIONEA=
	\left\{\begin{array}{lllllll}
		[\RDESCRIZIONEA{0}\equiv \langle \emptyset, 0, \DISTA, 0, 0\rangle ] 
		& \text{ if } \PRIZE{\ROOT}<0\\
		\text{[}\RDESCRIZIONEA{0}\equiv \langle \emptyset, 0, \DISTA-\PRIZE{\ROOT}, 0, 0\rangle, 
		\RDESCRIZIONEA{1}\equiv \langle \{\ROOT\}, \DISTA-\PRIZE{\ROOT}, \DISTA, \PROOTA, \PRIZE{\ROOT}\PROOTA\rangle ]
		& \text{ if } \PRIZE{\ROOT}\in (0,\DISTA)\\
		\text{[}\RDESCRIZIONEA{0}\equiv \langle \{\ROOT[A]\}, 0, \DISTA, \PROOTA, \PRIZE{\ROOT}\PROOTA\rangle ]
		& \text{ if } \PRIZE{\ROOT}\geq\DISTA
	\end{array}\right.
\end{align*}

We only highlight two special cases.
The case $\DISTA=0$ reduces the domain of  \FCA\ to a single point interval $[0,0]$.
On the other hand, having $\PRIZE{\ROOT}<0$ leads to only one optimal set (the empty one) on the whole domain, excluding a priori node \ROOT\ from all the possible \OPTIMALSETS\ independently of its probability. 
{As mentioned at the beginning of Section \ref{SEC:notation}, this} is particularly useful for modeling nodes representing just a bifurcation in the network and not a customer.

\subsubsection{The general case}
Unfortunately, we have no guarantee that for a certain $x\in[0,\DISTA]$ the \OPTIMALSET\ for \FCA\ is the union of the 
\OPTIMALSETS\ corresponding to the same  $x$ for \FC{{\SUBTREE{A}}} of the sub-trees \SUBTREE{A} with $i=0,1,\ldots,m$.
According to \eqref{eq:prop_Profitto_su_operazioni_insiemi} we know that the union of two \OPTIMALSETS\ $S_i$ and $S_j$ for the \SUBTREECHARFUNCTIONS\ \FC{{\SUBTREE[i]{A}}} and \FC{{\SUBTREE[j]{A}}} of distinct sub-trees in \TREE\ may lead to an expected profit larger than the sum of two profits alone, and the optimal profit may be even larger. In a formula
$$
\FC{{\SUBTREE[i]{A}}}+\FC{{\SUBTREE[j]{A}}}=\MGUADAGNOB{S_i}{x}+\MGUADAGNOB{S_j}{x}\leq\MGUADAGNOB{S_i\cup S_j}{x}\leq \FCSETINDEXA{\{i,j\}}\leq\FCA.
$$
Based on monotony of the \TREECHARFUNCTION\ and the matryoshka property of its \MAXOPTIMALSETS, intuition tells us that the selection of the \OPTIMALSET\ for \FC{{\SUBTREE[i]{A}}} encourages the selection of a node set in \SUBTREE[j]{A} even larger of the \OPTIMALSET\ for \FC{{\SUBTREE[j]{A}}} because the cost to reach nodes in \SUBTREE[j]{A} is partially paid by the engagement on \OPTIMALSET\ in sub-tree \SUBTREE[i]{A}.

Proposition \ref{P:profitto_con_virtual_bonus} confirms, quantifies and generalizes the intuition providing a technical basis for the proof of the main statement in Proposition \ref{P:FC_matrAlbero_contenuta_ProdottoCartesianoSottoalberi_BIS} stating that the \MAXOPTIMALSETS\ for \FCA\ have to be taken within the Cartesian product of the sub-tree matryoshkas.

\begin{proposition}\label{P:profitto_con_virtual_bonus}
	Given a tree \TREEA, let $I',I''\subseteq\TREEINDEX$ be two disjoint groups of sub-trees (i.e. $I'\cap I''=\emptyset$) and let $\Psi\subseteq\SETAINDEXED[I']$ and $\Phi\subseteq\SETAINDEXED[I'']$ two subsets of nodes.
	Then for any $x\in[0,\DISTA]$ we have 
	\begin{align}
		\MGUADAGNOB{\Psi\cup\Phi}{x} &= \MGUADAGNOB{\Psi}{x} + \MGUADAGNOB{\Phi}{\VBENZ{\Psi}}\label{eq:profitto_con_virtual_bonus}
	\end{align}
	where $\PSETS{\Psi}=0$ in case $\Psi=\emptyset$.
\end{proposition}

\begin{proof}[Proof of Proposition \ref{P:profitto_con_virtual_bonus}]
Result can be proved directly by the two following identities: 

	\begin{align*}
		\MRICAVO{\Psi\cup\Phi} &= \MRICAVO{\Psi} + \MRICAVO{\Phi}\\
		\MCOSTOB{\Psi\cup\Phi}{x} &= \MCOSTOB{\Psi}{x} + \MCOSTOB{\Phi}{\VBENZ{\Psi}}
	\end{align*}
	
	The first identity can be easily verified since $\Psi\cap\Phi=\emptyset$.
	Regarding the second identity, we first observe that $\VBENZ{\Psi}\leq\DISTA$, thus $\MCOSTOB{\Phi}{\VBENZ{\Psi}}$ is well defined being  $\Phi\subseteq\SET$.
	
	Since $\Psi\cup\Phi\subseteq\SET$, the identity can be proved by decomposing   $\MCOSTOB{\Psi\cup\Phi}{x}$ as shown in   \eqref{eq:costo_atteso_con_sviluppo_bonus} and \eqref{eq:costo_atteso_con_sviluppo_sottoalberi}
	\begin{align*}
		\MCOSTOB{\Psi\cup\Phi}{x}
		&= (\DISTA-x)\PSETS{\Psi\cup\Phi}+\SUMTREE\MCOSTOB{(\Psi\cup\Phi)\cap\SUBSET{A}}{\DISTA}\\
		&= (\DISTA-x)\PSETS{\Psi\cup\Phi}+\sum_{i\in I'}\MCOSTOB{\Psi\cap\SUBSET{A}}{\DISTA}+\sum_{i\in I''}\MCOSTOB{\Phi\cap\SUBSET{A}}{\DISTA}+\sum_{i\in\TREEINDEX\setminus(I'\cup I'')}\MCOSTOB{\emptyset\cap\SUBSET{A}}{\DISTA}\\
		&= (\DISTA-x)\PSETS{S}+\MCOSTOB{\Psi}{\DISTA}+\MCOSTOB{\Phi}{\DISTA}\\
		&= (\DISTA-x)\PSETS{S}+\left[(\DISTA-x)\PSETS{\Psi}+\MCOSTOB{\Psi}{\DISTA}\right]-(\DISTA-x)\PSETS{\Psi}+\MCOSTOB{\Phi}{\DISTA}\\
		&=\MCOSTOB{\Psi}{x}+(\DISTA-x)\PSETS{S}-(\DISTA-x)\PSETS{\Psi} + \MCOSTOB{\Phi}{\DISTA}\\
		&=\MCOSTOB{\Psi}{x}+(\DISTA-x)(\PSETS{S}-\PSETS{\Psi}) + \MCOSTOB{\Phi}{\DISTA},
	\end{align*}

	and then, % $\MCOSTOB{\Phi}{\VBENZ{\Psi}}$ 
	according to \eqref{eq:costo_atteso_con_sviluppo_bonus},
	\begin{align*}
		\MCOSTOB{\Phi}{\VBENZ{\Psi}}
		&= (\DISTA- (\VBENZ{\Psi}) )\PSETS{\Phi}+\MCOSTOB{\Phi}{\DISTA}\\
		&= (\DISTA-x)(\PCOMP{\Psi})\PSETS{\Phi}+\MCOSTOB{\Phi}{\DISTA}.
	\end{align*}
	
	Now, it is enough to show that the following equality holds
	$$
	(\DISTA-x)(\PSETS{S}-\PSETS{\Psi}) = (\DISTA-x)(\PCOMP{\Psi})\PSETS{\Phi}
	$$
	that, through algebraic computations, corresponds to the identity
	$$
	\PSETS{S}= \PSETS{\Psi\cup\Phi}= \PSETS{\Phi}+\PSETS{\Psi}- \PSETS{\Psi}\PSETS{\Phi}.
	$$
\end{proof}

\begin{proposition}\label{P:FC_matrAlbero_contenuta_ProdottoCartesianoSottoalberi_BIS}
	Let  $I\subseteq\TREEINDEX$ be a group of sub-trees and let $I=I'\cup I''$ with $I'\cap I''=\emptyset$ an arbitrary partition of $I$. For each set of nodes $S\in\MATRIOSKAAI$ let $S'=S\cap\SETAINDEXED[I']$ and $S''=S\cap\SETAINDEXED[I'']$ be its components in the two groups of sub-trees (i.e.\ $S=S'\cup S''$ and $S'\cap S''=\emptyset$).
	
	Then the following condition holds:
	$$
	(S',S'') \in \MATRIOSKAAI[I']\bigtimes\MATRIOSKAAI[I'']
	$$
\end{proposition}

\begin{proof}[Proof of proposition \ref{P:FC_matrAlbero_contenuta_ProdottoCartesianoSottoalberi_BIS}]
	Let $\hat{x}\in[0,\DISTA]$ be a value of the bonus for which  $S$ is a  \MAXOPTIMALSET\ of the \SUBTREECHARFUNCTION\ \FCSETINDEXA{I}. 
	
	We first prove that $S'\in\MATRIOSKAAI[I']$ by contradiction. 
	From  Proposition \ref{P:profitto_con_virtual_bonus} we know that
	$$
	\MGUADAGNOB{S}{\hat{x}} = \MGUADAGNOB{S''}{\hat{x}} + \MGUADAGNOB{S'}{\VBENZ[\hat{x}]{S''}},
	$$%
	if $S'\notin\MATRIOSKAAI[I']$ then $S'$ cannot be a \MAXOPTIMALSET\ for any $x\in[0,\DISTA]$.
	If $S'$ is not optimal  for any value of the bonus, then it is not even for
	$x=\VBENZ[\hat{x}]{S''}$, and thus $\MGUADAGNOB{S}{\hat{x}}$ can be strictly increased which contradicts the optimality of $S$. 
	If  $S'$ is optimal for  $x=\VBENZ[\hat{x}]{S''}$, but not maximal, then also  $S$ can not be a \MAXOPTIMALSET. 
	Thus, $S'$ is necessarily a \MAXOPTIMALSET\ for \FCSETINDEXA{I'} for at least one value in $[0,\DISTA]$ and thus it belongs to  \MATRIOSKAAI[I'] by definition. 
	
	A similar discussion proves that $S''\in\MATRIOSKAAI[I'']$.
\end{proof}

\medskip
It is worth noticing  that the Cartesian product $\MATRIOSKAAI[I']\bigtimes\MATRIOSKAAI[I'']$
of two matryoshkas	produces a set of pairs $(S',S'')$ of node sets 
with $S'\in\MATRIOSKAAI[I']$ and $S''\in\MATRIOSKAAI[I'']$, whereas the matryoshka of the global group contains sets such as $(S'\cup S'')$. From now on, for sake of simplicity, 
we will assimilate, with a little abuse of notation, 
$(S',S'')$ and $(S'\cup S'')$ and use, as a consequence, expression as 
$$\MATRIOSKAAI[I'\cup I''] \subset \MATRIOSKAAI[I']\bigtimes\MATRIOSKAAI[I''].$$

By applying several times the property shown in Proposition \ref{P:FC_matrAlbero_contenuta_ProdottoCartesianoSottoalberi_BIS} and 
by exploiting the associative property of the Cartesian product, since we are interested only in the set union of the nodes involved, we can easily observe that $$\MATRIOSKAA\subseteq\CARTMATRIOSKA$$
with any order of product execution; hence, once known the matryoshkas \MATRIOSKA{\SUBTREE{A}}, each one with  $O(|\SUBSET{A}|)$ optimal sets, the computation of \MATRIOSKAA\ does not require the evaluation of $O(2^{|\SET|})$ \SETCHARFUNCTIONS\ anymore, but in the worst case of  $O(\prod_{i=0}^m|\SUBSET{A}|)=O(|\SET|^{m+1})$. 

\medskip
It is now appropriate to specify that, while matryoshka \MATRIOSKAA,
which we are trying to build, describes \FCA\ in the interval $[0,\DISTA]$,  matryoshkas  \MATRIOSKA{\SUBTREE{A}} describe functions  \FCSA\ in larger intervals $[0,\DIST{\SUBROOT[i]{A}}]$ and might contain more elements than necessary. 
Proposition \ref{P:LIMTE_MATRIOSKE}
will help us to adequately limit matryoshkas extension for each  sub-tree from which \MATRIOSKAA\ will be obtained. Their size remain, however, in the order of $O(|\SUBSET{A}|)$.

\begin{proposition}\label{P:LIMTE_MATRIOSKE}
	Let us consider a tree \TREEA.
	The last  \MAXOPTIMALSET\ of the matryoshka \MATRIOSKAA\ is the set union of \MAXOPTIMALSETS\ for \FC[\DISTA]{\SUBTREE{A}}.
\end{proposition}
\begin{proof}[Proof of Proposition \ref{P:LIMTE_MATRIOSKE}]

	From expression \eqref{eq:guadagno_atteso_con_sviluppo_sottoalberi} we know that for any node set $S\subseteq\SET$ and $x=\DISTA$ we have
	$$
	\MGUADAGNOB{S}{\DISTA}=\SUMTREE\MGUADAGNOB{S\cap\SUBSET{A}}{\DISTA}.
	$$
	Thus, maximization of $\MGUADAGNOB{S}{\DISTA}$ with respect to $S\subseteq\SET$ is equivalent to solve $m+1$ independent problems on the $m+1$ sub-trees of \TREE.
\end{proof}

\subsubsection{Order of magnitude of {matryoshka's Cartesian products}}\label{SEC:ordine_prodotto_cartesiano}
Computing the whole Cartesian Product of all matryoshkas associated to sub-trees may be not computationally convenient. 
In fact, if $m=O(|\SET|)$ also the Cartesian product contains $O(2^{|\SET|})$ elements. 
Consider the case of a simple tree with $m$ nodes distributed in $m$ sub-trees of one node each (including the root sub-tree \SUBTREE[0]{A}); 
the matryoshka of each sub-tree potentially contains  $2$ \OPTIMALSETS\ and their Cartesian product has cardinality $O(2^m)$.

The situation changes radically if, to obtain the matryoshka \MATRIOSKAA, we first calculate the matryoshkas of two complementary non-empty sub-trees groups $I'\subset\TREEINDEX$ and $I''=\TREEINDEX\setminus I'$. 
In this case, the two matryoshkas \MATRIOSKAAI[I'] and \MATRIOSKAAI[I''] contain a number $O(|\SET|)$  of elements, but the cardinality of their Cartesian product does not go beyond $O(|\SET|^2)$.
A sequential grouping strategy allows to complete the grouping process in $O(|\TREEINDEX|)$ iterations, each of which involves a Cartesian product of dimension $O(|\SET|^2)$.
At this point we have a strong clue that the computation of \DESCRIZIONEA\ for \FCA\ can be done in polynomial time. 
However, before drawing conclusions, let's make a few more observations that will allow us to further lower the degree of the polynomial.

\subsubsection{Avoiding the Cartesian product of matryoshkas}\label{SEC:evitare_prodotto_cartesiano}

We have shown that the matryoshka \MATRIOSKAA\ is contained in the Cartesian product of two matryoshkas each of  $O(|\SET|)$ elements. 
Therefore, the product has cardinality $O(|\SET|^2)$. 
In this section we show that it is not necessary to evaluate all the $O(|\SET|^2)$ elements of the Cartesian product and that we can instead select in $O(|\SET|)$ time a matryoshka \MATRIOSKAK\  which certainly contains  \MATRIOSKAA. As a final step we will see that we can get \MATRIOSKAA\ from \MATRIOSKAK\ always in $O(|\SET|)$ time. 
This will be the decisive step that allows us to define a quadratic algorithm to compute \FCA\ and solve the problem.

Thus, we will discuss how, the matryoshkas 
$$
\MATRIOSKAB=\left[ \MATSETB{0} \subset \MATSETB{1} \subset \ldots \subset \MATSETB{\CARDB}\right]
$$
and
$$
\MATRIOSKAC=\left[ \MATSETC{0} \subset \MATSETC{1} \subset \ldots \subset \MATSETC{\CARDC}\right]
$$
of two disjoint groups $\GRUPPOB\subset\TREEINDEX$ and $\GRUPPOC\subset\TREEINDEX$ of sub-trees %, the corresponding matryoshkas \MATRIOSKAAI[\GRUPPOB] and \MATRIOSKAAI[\GRUPPOC] 
can be processed to give rise to the matryoshka 
$$
\MATRIOSKAD=\left[ \MATSETD{0} \subset \MATSETD{1} \subset \ldots \subset \MATSETD{\CARDD}\right]
$$
of the union group $\GRUPPOD=\GRUPPOUBC$. 
The whole operation can be done in time $O(\CARDB+\CARDC)$.
On the basis of the Proposition  \ref{P:LIMTE_MATRIOSKE}, we assume that \MATRIOSKAB\ and \MATRIOSKAC\ cover only the domain $[0,\DISTA]$.

\medskip
First we will identify %in time $O(\CARDB+\CARDC)$ 
a matryoshka $\MATRIOSKAK\subset\CARTMATRIOSKABIN$ such that $\MATRIOSKAD\subseteq\MATRIOSKAK$. 
Then we will see how to extract %, always in time $O(\CARDB+\CARDC)$, 
the matryoshka \MATRIOSKAD.

\smallskip
The idea is to start with a matryoshka 
%\replace{$\MATRIOSKAKH[0]=[\MATSETK{0}]$ whose only element, being the union of the two minimal sets in the respective matryoshkas (i.e. $\MATSETK{0}=\MATSETB{0}\cup \MATSETC{0}$), }
{$\MATRIOSKAKH[0]=[\MATSETK{0}=\MATSETB{0}\cup \MATSETC{0}]$ whose only element}
is clearly a subset of $\MATSETD{0}\in\MATRIOSKAD\subset\CARTMATRIOSKABIN$.

Then, for $h=1,2,\ldots,\CARDB+\CARDC$, a set $\MATSETK{h}\in\CARTMATRIOSKABIN$ is iteratively added to \MATRIOSKAKH[h-1] so that a new matryoshka \MATRIOSKAKH[h] compliant with the following definition of coherence  with \MATRIOSKAD\ is obtained at each step.

\begin{definition}
	A matryoshka  $\MATRIOSKAKH=[\MATSETK{0} \subset \MATSETK{1} \subset \ldots \subset \MATSETK{h}]$ is said to be  \textbf{coherent} with a matryoshka  \MATRIOSKAD\ if for any $\MATSETD{}\in\MATRIOSKAD$ it holds that 
		$$
		\MATSETD{}\in \MATRIOSKAKH\quad \vee\quad \MATSETK{h}\subseteq \MATSETD{}.
		$$
\end{definition}

The key point of the definition  is that if there exists a node set $\MATSETD{}\in\MATRIOSKAD$ not belonging to  \MATRIOSKAKH, then $\MATSETD{}$ must contain the biggest element of \MATRIOSKAKH, which guarantees the possibility of extending the matryoshka \MATRIOSKAKH\ reaching first the smallest of such sets and from that all the others.

The question now is how to extend \MATRIOSKAKH. 
We know that $\MATSETK{h}=\MATSETB{i}\cup\MATSETC{j}\in\CARTMATRIOSKABIN$ for some $i\leq\CARDB$ and $j\leq\CARDC$.
Let us assume that $i+j<\CARDB+\CARDC$, otherwise we already have $\MATSETD{\CARDD}\subseteq\MATSETK{h}$ and there is no need to extend matryoshka \MATRIOSKAKH.
The smallest step we can do is to take either $\MATSETK{h+1}=\MATSETB{i+1}\cup\MATSETC{j}$ or $\MATSETK{h+1}=\MATSETB{i}\cup\MATSETC{j+1}$. 
Both choices may be correct, but in some cases only one preserves the property of coherence with \MATRIOSKAD.
For example, consider matryoshka $\MATRIOSKAKH[0]=[\MATSETK{0}]$ with $\MATSETK{0}=\MATSETB{0}\cup \MATSETC{0}$ and its candidate extensions $Y=\MATSETB{1}\cup \MATSETC{0}$ and $Z=\MATSETB{0}\cup \MATSETC{1}$.
In the case $\MATSETD{0}=\MATSETB{2}\cup\MATSETC{3}$, then both $Y$ and $Z$ are valid extensions that preserve coherence with \MATRIOSKAD, but in the case $\MATSETD{0}=\MATSETB{0}\cup\MATSETC{3}$, only $Z$ is a correct choice as $Y$ would irremediably broke the coherence of the new matryoshka with \MATRIOSKAD.

Thus, we need a rule that prevent us from taking a wrong choice.
Proposition \ref{P:MTK_da_BiCj_espandi_matrioska_coerente} gives us the rule we need. 
The idea exploits equation \eqref{eq:profitto_con_virtual_bonus} in Proposition \ref{P:profitto_con_virtual_bonus}. 
Let us consider a bonus $x<\DISTA$, two sub-tree groups \GRUPPOB\ and \GRUPPOC\  and the corresponding matryoshkas $\MATRIOSKAB$ and $\MATRIOSKAC$.
A commitment to serve node set $\MATSETB{i}\in\MATRIOSKAB$ with a bonus $x$ offers a larger bonus $\VBENZ{\MATSETB{i}}$ to visit nodes in the sub-tree group \GRUPPOC.
Which set in \MATRIOSKAC\ is better to select (given the commitment to \MATSETB{i}) depends on the bonus $x$.
Thus, we define the \textit{entry threshold} of set $\MATSETC{j}$ given the commitment to \MATSETB{i} as the minimum bonus value such that the choice of \MATSETC{j} is better than any other (given the commitment to \MATSETB{i}).
Of course the same considerations can be made on the other side, by considering a commitment to some $\MATSETC{j}\in\MATRIOSKAC$ and subsequent choice of an \OPTIMALSET\ in $\MATRIOSKAB$.
The following definition helps to formalize the concept of entry threshold and Proposition \ref{P:SOGLIE} states the main properties of these indices that will be used in Proposition \ref{P:MTK_da_BiCj_espandi_matrioska_coerente}.

\begin{definition} 
    Let \GRUPPOB\ and \GRUPPOC\ be two disjoint groups of sub-trees in tree \TREE\ (i.e. $\GRUPPOB, \GRUPPOC\subseteq\TREEINDEX$, $\GRUPPOB\cap\GRUPPOC=\emptyset$), and
	let $\MATSETB{i}\in\MATRIOSKAB$ and $\MATSETC{j}\in\MATRIOSKAC$.
	
	We define 
	\textit{\SOGLIAINGRESSO} of the 'entering' node set $\MATSETC{j}$ given the committed node set  $\MATSETB{i}$ the value
	$$
	\SOGLIA{\MATSETC{j}}{\MATSETB{i}}=\underset{x}{\arg\min}\left\{\MGUADAGNOB{\MATSETB{i}\cup \MATSETC{j}}{x}=\max_{h}\MGUADAGNOB{\MATSETB{i}\cup \MATSETC{h}}{x}    \right\}
	$$
	
	Symmetrically, we define \textit{\SOGLIAINGRESSO} of the entering node set $\MATSETB{i}$ given the committed node set $\MATSETC{j}$ the value
	$$
	\SOGLIA{\MATSETB{i}}{\MATSETC{j}}=\underset{x}{\arg\min}\left\{\MGUADAGNOB{\MATSETB{i}\cup \MATSETC{j}}{x}=\max_{h}\MGUADAGNOB{\MATSETB{h}\cup \MATSETC{j}}{x}    \right\}
	$$
	
\end{definition}

\begin{proposition}\label{P:SOGLIE}
    Let \GRUPPOB\ and \GRUPPOC\ be two disjoint groups of sub-trees in tree \TREE, and
	let $\MATRIOSKAB=[\MATSETB{0},\ldots,\MATSETB{\CARDB}]$ and $\MATRIOSKAC=[\MATSETC{0},\ldots,\MATSETC{\CARDC}]$ be the respective matryoshkas of \MAXOPTIMALSETS.
	For each $i\leq\CARDB$, $j\leq\CARDC$ we have:
	\begin{align}
		&    \SOGLIA{\MATSETC{j}}{\MATSETB{i}}=\REQUIREREC{\GRUPPOC}{j}{\MATSETB{i}}\leq \RDXMINBC{\GRUPPOC}{j}             & \text{{entry advance}}\label{eq:SOGLIA_CB_espressione}\\
		&    \SOGLIA{\MATSETC{1}}{\MATSETB{i}}<\SOGLIA{\MATSETC{2}}{\MATSETB{i}}<\ldots<\SOGLIA{\MATSETC{\CARDC}}{\MATSETB{i}}     & \text{{monotonicity with respect to the entering set}}\label{eq:SOGLIA_CB_monotonia_C}\\
		&    \SOGLIA{\MATSETC{j}}{\MATSETB{0}}>\SOGLIA{\MATSETC{j}}{\MATSETB{1}}>\ldots>\SOGLIA{\MATSETC{j}}{\MATSETB{\CARDB}} & \text{{monotonicity with respect to the committed set}}\label{eq:SOGLIA_CB_monotonia_B}
	\end{align}
	Obviously the symmetrical ones also apply
	\begin{align}
		&    \SOGLIA{\MATSETB{i}}{\MATSETC{j}}=\REQUIREREC{\GRUPPOB}{i}{\MATSETC{j}}\leq \RDXMINBC{\GRUPPOB}{i}      \label{eq:SOGLIA_BC_espressione}\\
		&    \SOGLIA{\MATSETB{1}}{\MATSETC{j}}<\SOGLIA{\MATSETB{2}}{\MATSETC{j}}<\ldots<\SOGLIA{\MATSETB{\CARDB}}{\MATSETC{j}}     \label{eq:SOGLIA_BC_monotonia_B}\\
		&    \SOGLIA{\MATSETB{i}}{\MATSETC{0}}>\SOGLIA{\MATSETB{i}}{\MATSETC{1}}>\ldots>\SOGLIA{\MATSETB{i}}{\MATSETC{\CARDC}}     \label{eq:SOGLIA_BC_monotonia_C}
	\end{align}
	
\end{proposition}

The proof of Proposition \ref{P:SOGLIE} is provided in Appendix. Here,
we observe that
inequality \eqref{eq:SOGLIA_CB_espressione} shows that node set $\MATSETC{j}\in\MATRIOSKAC$ starts to dominate smaller sets in its matryoshka for smaller values of the bonus when also set $\MATSETB{i}\in\MATRIOSKAB$ is committed.
Moreover, the inequality chain \eqref{eq:SOGLIA_CB_monotonia_C} attests that the relative order of set selection in \MATRIOSKAC\ with respect to the bonus remains unchanged when set $\MATSETB{i}\in\MATRIOSKAB$ is committed.
Finally, the inequality chain \eqref{eq:SOGLIA_CB_monotonia_B} shows that the entry threshold of \MATSETC{j} becomes smaller and smaller as the set committed in \MATRIOSKAB\ becomes larger.
Proposition \ref{P:MTK_da_BiCj_espandi_matrioska_coerente} will use these facts to define a rule to coherently extend a matryoshka. The proof is provided in the Appendix.

\begin{proposition}\label{P:MTK_da_BiCj_espandi_matrioska_coerente}
    Let \GRUPPOB\ and \GRUPPOC\ be two disjoint groups of sub-trees in tree \TREE, 
	let the matryoshka $\MATRIOSKAKH=$ $\left[\MATSETK{0} \subset \MATSETK{1}\subset \ldots \subset \MATSETK{h}\right]\subseteq\CARTMATRIOSKABIN$ be coherent with \MATRIOSKAUBC, 
	and let $\MATSETK{h}=\MATSETB{i}\cup \MATSETC{j}$ denote its last element with $i+j<\CARDB+\CARDC$. 
	Then, by choosing $\MATSETK{h+1}$ with the following rule
	\begin{itemize}
		\item If $(i<\CARDB,\ j=\CARDC) \Rightarrow \MATSETK{h+1}=\MATSETB{i+1}\cup \MATSETC{j}$
		\item Otherwise if $(i=\CARDB,\ j<\CARDC) \Rightarrow \MATSETK{h+1}=\MATSETB{i}\cup \MATSETC{j+1}$
		\item Otherwise if $(\SOGLIA{\MATSETB{i+1}}{\MATSETC{j}}\leq\SOGLIA{\MATSETC{j+1}}{\MATSETB{i}}) \Rightarrow \MATSETK{h+1}=\MATSETB{i+1}\cup \MATSETC{j}$
		\item Otherwise  $[\SOGLIA{\MATSETB{i+1}}{\MATSETC{j}}\geq\SOGLIA{\MATSETC{j+1}}{\MATSETB{i}}] \Rightarrow \MATSETK{h+1}=\MATSETB{i}\cup \MATSETC{j+1}$
	\end{itemize}
the extension of \MATRIOSKAKH\ with $\MATSETK{h+1}$ produces a matryoshka \MATRIOSKAKH[h+1] coherent with \MATRIOSKAUBC.
\end{proposition}

In conclusion of this Section we observe that so far we have seen how to 
build the matryoshka of a leaf node  and commented that the matryoshkas of two (groups of) sub-trees can be "merged" to form ever larger groups of sub-trees up to form the matryoshka of the father tree.
We also showed that a matryoshka containing the matryoshka of the union of two (groups of) sub-trees can be computed in a number of steps which is linear with respect to the number on nodes involved.
How to assemble all this stuff and obtain an efficient algorithm is described in Section \ref{SEC:algorithm} where we provide an algorithm definition and discuss its time complexity.

\section{An optimal algorithm}\label{SEC:algorithm}

In this section, we formalize the algorithm and discuss its computational complexity.
As a first step we discuss the fundamental step of creating the description of the \SUBTREECHARFUNCTION\ of a group of sub-trees which is the union of two disjointed groups of sub-trees of which we have the descriptions (Algorithm \texttt{Merge}).
Then, we provide the definition of a recursive procedure that cumulatively merge the descriptions of all sub-trees in a given tree (Algorithm \texttt{SolveTree}), and finally the main function providing the required problem solution (Algorithm \texttt{PPTP-Tree}).

\begin{proposition}\label{P:cpl_algo_Merge}
Let \TREE\ be a tree and $\GRUPPOB\in\TREEINDEX$, $\GRUPPOC\in\TREEINDEX$ be two disjointed groups of sub-trees. 
The description \DESCRIZIONEAI[\GRUPPOB\cup\GRUPPOC] of the \SUBTREECHARFUNCTION\ of the union group $\GRUPPOB\cup\GRUPPOC$ can be computed in time $O(|\SETAINDEXED[(\GRUPPOB\cup\GRUPPOC)]|)$ starting from \DESCRIZIONEAB and \DESCRIZIONEAC.
%	Given the descriptions of the \SUBTREECHARFUNCTIONS\ of two groups of sub-trees \GRUPPOB\ and \GRUPPOC\ of sub-trees of a tree
%	\TREEA\ having $\CARDB$ and $\CARDC$ nodes each, the description
%	of \SUBTREECHARFUNCTION\ of the union group $\GRUPPOB\cup\GRUPPOC$ can be computed in time $O(\CARDB+\CARDC)$.
\end{proposition}

\begin{algorithm}[H]\label{A:Merge}
	\caption{\texttt{Merge}}
	
	\DontPrintSemicolon
	\SetKwInOut{Input}{input}\SetKwInOut{Output}{output}\SetKwInOut{Global}{global}
	\Input{\DESCRIZIONEAB, \DESCRIZIONEAC\ descriptions of \FCSETINDEXA{\GRUPPOB} and \FCSETINDEXA{\GRUPPOC}} %\\ \DISTA, length of common path}
	\Output{\DESCRIZIONEAUBC\ description of \FCSETINDEXA{\GRUPPOUBC}}%\ in $[0,\DISTA]$}
	
	\smallskip	
	-- $\DISTA=\ENVDIS{\GRUPPOB}{\CARDB}{\FIELDXMAX}$\;
	-- $E = $ empty stack\;
	-- $K.\FIELDSET = \ENVDIS{\GRUPPOB}{0}{\FIELDSET}\cup\ENVDIS{\GRUPPOC}{0}{\FIELDSET}$\;\label{step:A3-a}
	-- $K.\FIELDPENDENZA = 1-(1-\ENVDIS{\GRUPPOB}{0}{\FIELDPENDENZA})(1-\ENVDIS{\GRUPPOC}{0}{\FIELDPENDENZA})$\;
	-- $K.\FIELDQZERO = \ENVDIS{\GRUPPOB}{0}{\FIELDQZERO} + \ENVDIS{\GRUPPOC}{0}{\FIELDQZERO}$ \tcp*[h]{value of $\MGUADAGNOB{K.\FIELDSET}{\DISTA}$}\; 
	-- $K.\FIELDXMIN = 0$, $K.\FIELDXMAX = \DISTA$\;\label{step:A3-b}
	-- $E.push(K)$\;
	
	\smallskip	
	-- $i=0$, $j=0$\;
	\While{$(i<\CARDB\ \vee\ j<\CARDC)$}{
		\lIf{$(i==\CARDB)$}{$j=j+1$}\label{step:A3-c}
		\lElseIf{$(j==\CARDC)$}{$i=i+1$}
		\Else{
			-- $S_B = \ENVDIS{\GRUPPOB}{i+1}{\FIELDSET}\cup\ENVDIS{\GRUPPOC}{j}{\FIELDSET}$\;
			-- $S_C = \ENVDIS{\GRUPPOB}{i}{\FIELDSET}\cup\ENVDIS{\GRUPPOC}{j+1}{\FIELDSET}$\;
			\lIf{$(\SOGLIA{S_B}{K.\FIELDSET}<\SOGLIA{S_C}{K.\FIELDSET})$}{$i=i+1$}\label{step:A3-S}
			\lElse{$j=j+1$}
		}\label{step:A3-d}
		-- $K'.\FIELDSET = \ENVDIS{\GRUPPOB}{i}{\FIELDSET}\cup\ENVDIS{\GRUPPOC}{j}{\FIELDSET}$\;\label{step:A3-e}
		-- $K'.\FIELDPENDENZA = 1-(1-\ENVDIS{\GRUPPOB}{i}{\FIELDPENDENZA})(1-\ENVDIS{\GRUPPOC}{j}{\FIELDPENDENZA})$\;
		-- $K'.\FIELDQZERO = \ENVDIS{\GRUPPOB}{i}{\FIELDQZERO} + \ENVDIS{\GRUPPOC}{j}{\FIELDQZERO}$ \tcp*[h]{value of $\MGUADAGNOB{K'.\FIELDSET}{\DISTA}$}\;
		-- $K'.\FIELDXMIN = 0$, $K.\FIELDXMAX = \DISTA$\;\label{step:A3-f}
		
		\smallskip	
		-- $x=\DISTA-\frac{K'.\FIELDQZERO-K.\FIELDQZERO}{K'.\FIELDPENDENZA-K.\FIELDPENDENZA}$ \tcp*[h]{Solution of equation $\MGUADAGNOB{K'.\FIELDSET}{x}=\MGUADAGNOB{K.\FIELDSET}{x}$}\;\label{step:A3-g}
		\While{$(x\leq K.\FIELDXMIN)$}{
			-- $E.pop()$\;
			-- $K=E.peek()$\;
			-- $x=\DISTA-\frac{K'.\FIELDQZERO-K.\FIELDQZERO}{K'.\FIELDPENDENZA-K.\FIELDPENDENZA}$\;
		}\label{step:A3-h}
		-- $K.\FIELDXMAX=x$\;\label{step:A3-i}
		-- $K'.\FIELDXMIN=x$\;\label{step:A3-j}
		-- $E.push(K')$\;
		-- $K = K'$\;
	}
	-- $\DESCRIZIONEAI[\GRUPPOUBC]=$ revert $E$\label{step:A3-k}
\end{algorithm}

\begin{proof}[Proof of Proposition \ref{P:cpl_algo_Merge}]
We refer to Algorithm \ref{A:Merge} (\texttt{Merge}) for the pseudo-code description  of the procedure that calculates the result.

\medskip
\textbf{Correctness.}
The procedure is based on Proposition \ref{P:MTK_da_BiCj_espandi_matrioska_coerente} and starts by building the smallest matryoshka \MATRIOSKAKH[0] coherent with
$\MATRIOSKAAI[\GRUPPOB\cup\GRUPPOC]$ using the smallest 
\MAXOPTIMALSETS\ of the matryoshkas in the descriptions
\DESCRIZIONEAB\ and \DESCRIZIONEAC\ of the \SUBTREECHARFUNCTIONS\ of the groups \GRUPPOB\ and \GRUPPOC (lines \ref{step:A3-a}-\ref{step:A3-b}).
The corresponding record descriptor $K$ is then added to the stack $E$ which represents the under construction description \DESCRIZIONEAUBC\ of \FC{\GRUPPOB\cup\GRUPPOC}.
In particular, the calculation of the field $q$ comes from the formula \eqref{eq:guadagno_atteso_con_sviluppo_sottoalberi}, and being the only set currently considered, its domain is over the whole interval $[0,\DISTA]$.

The selection rule described in Proposition \ref{P:MTK_da_BiCj_espandi_matrioska_coerente} (lines \ref{step:A3-c}-\ref{step:A3-d}) is applied to each iteration of the main cycle to extend the matryoshka \MATRIOSKAKH\ to \MATRIOSKAKH[h+1] and generate a new record $K'$ to add to the stack $E$ (lines \ref{step:A3-e}-\ref{step:A3-f}). 

Note that the matryoshka coherent with $\MATRIOSKAAI[\GRUPPOB\cup\GRUPPOC]$, although calculated element by element in the variable $K'$, is not explicitly stored as the relative records are immediately used to build the envelope of \FC{\GRUPPOB\cup\GRUPPOC} in the stack $E$.
Indeed, node set $K'.S$ certainly has the right to be included in $E$ since, by virtue of its being a superset of all the sets inserted up to now in $E$, it turns out to have an expected profit not smaller than all these at least for $x=\DISTA$ (see equation \eqref{eq:guadagno_atteso_con_sviluppo_sottoalberi}). 
Thus, before inserting $K'$ into $E$, the records referred to sets dominated, over their entire range, by $K'.\FIELDSET$  are removed from $E$ (lines \ref{step:A3-g}-\ref{step:A3-h}). 
In general, the first set that is not fully dominated will yield part of its interval to the new set $K'.S$, so an adjustment of the two ranges follows at the intersection of their respective \SETCHARFUNCTIONS\ (lines \ref{step:A3-i}-\ref{step:A3-j}).
Finally the stack $E$ is reverted to sort records of \DESCRIZIONEAI[\GRUPPOUBC] in the right order (line \ref{step:A3-k}). 

\medskip
\textbf{Complexity.}
The main loop is executed no more than $\CARDB + \CARDC\leq|\SETAINDEXED[\GRUPPOUBC]|$ times because at each iteration the index $i$ or the index $j$ is incremented by $1$.
Observe that, despite their unsettling definition, the \SOGLIEINGRESSO\ computed at line \ref{step:A3-S} can be computed in $O(1)$ time according to formulas \eqref{eq:SOGLIA_CB_espressione} and \eqref{eq:SOGLIA_BC_espressione} in Proposition \ref{P:SOGLIE}.
Thus, all operations in the loop have a cost of $O(1)$ except for the inner loop which removes a certain number of records from the stack $E$ before inserting the record for the new $K$.
Each iteration of the inner loop has a cost of $O(1)$, and, since at each iteration a record is removed from the stack $E$, the number of iterations is bounded by the number of records in the stack. 
The number of records remaining at loop termination determines the bound for the maximum number of iterations on the next pass. 
Basically, in the first pass we have $|E|=1$, and $h_1\leq1$ iterations are performed, for $2-h_1$ records in $E$ at the exit from the loop.
In the second pass, $h_2\leq2-h_1$ iterations are performed for $3-h_1-h_2$ records in $E$ when exiting the loop.
In the third pass, $h_3\leq3-h_1-h_2$ iteration are performed for $4-h_1-h_2-h_3$ records in $E$ when leaving the loop.
In the $k$-th pass, $h_k\leq k-\sum_{j<k}h_j$ iterations are performed for $k-\sum_{j\leq k}h_j$ records in $E$ upon exiting the loop.
%
%
%Basically, in the first pass $|E|=1$, maximum $h_1\leq1$ iterations, for 
%$2-h_1$ records in  $E$ at the exit from the loop;
%in the second pass, $h_2\leq2-h_1$ iterations for $3-h_1-h_2$
%records in $E$ when exiting the loop; in the third pass,
%
%$h_3\leq3-h_1-h_2$ iteration for $4-h_1-h_2-h_3$ records in $E$ when leaving the loop; 
%in the $k$-th pass, $h_k\leq k-\sum_{j<k}h_j$ iterations for $k-\sum_{j\leq k}h_j$ records in $E$ upon exiting the loop.
%
After $\CARDB+\CARDC$ steps, we get $\sum_{j\leq{\CARDB+\CARDC}}h_j\leq \CARDB+\CARDC$ 
which shows that, overall, the internal loop to extract records from the stack $E$ has a cost $O(\CARDB+\CARDC)$.
This ends the proof.
\end{proof}

\medskip
We can now state the complexity to find the optimal solution for the PPTP-T:

\begin{theorem}\label{T:main}
	An \OPTIMALSET\ for problem \eqref{DEF:PROBLEMA_STOCASTICO_BASE_BONUS} can be computed in $O(|\SETT|^2)$ time.
\end{theorem}

\medskip
\begin{algorithm}[H]\label{A:PPTP-Tree}
	\caption{\texttt{PPTP-Tree}}
	
	\DontPrintSemicolon
	\SetKwInOut{Input}{input}\SetKwInOut{Output}{output}\SetKwInOut{Global}{global}
	\Input{\TREET\ tree}
	\Output{$S$ optimal node set of \TREET; $G$ optimal value}
	
	-- $\DESCRIZIONE{\TREET}=\texttt{SolveTree(\TREET,\TREEINDEX[T])}$\tcp*[h]{description of \TREECHARFUNCTION\ \FC{\TREET}}\;\label{Astep:solve}
	-- $S=\RDTSET{0}$\;
	-- $G=\RDTQZERO{0} -\DISTT\cdot\RDTPENDENZA{0}$\;
\end{algorithm}

\bigskip

\begin{algorithm}[H]\label{A:SolveTree}
	\caption{\texttt{SolveTree}}
	
	\DontPrintSemicolon
	\SetKwInOut{Input}{input}\SetKwInOut{Output}{output}\SetKwInOut{Global}{global}
	\Input{\TREEA\ a tree}
	\Output{\DESCRIZIONEA\ description of \TREECHARFUNCTION\ \FCA}
	
	\smallskip	
	-- compute \DESCRIZIONE{\SUBTREE[0]{A}} \tcp*[l]{See base case in Section \ref{SEC:Computing_FCA}}
	-- $\DESCRIZIONEA=\texttt{Truncate(\DESCRIZIONE{\SUBTREE[0]{A}},\DISTA)}$ \tcp*[h]{truncate description at root of tree \TREEA}\;
	\For{$(i=1 \text{ to } |\TREEINDEX|-1)$}{
		-- $\DESCRIZIONE{\SUBTREE{A}} = \texttt{SolveTree($\SUBTREE[i]{A}$)}$\;
		-- $\DESCRIZIONE{\SUBTREE{A}} = \texttt{Truncate(\DESCRIZIONE{\SUBTREE{A}},\DISTA)}$\;
		-- $\DESCRIZIONEA=\texttt{Merge(\DESCRIZIONEA,\DESCRIZIONE{\SUBTREE{A}})}$\;
	}
\end{algorithm}

\begin{proof}[Proof of Theorem \ref{T:main}]
We refer to Algorithm  \ref{A:PPTP-Tree} (\texttt{PPTP-Tree}) which builds 
 \DESCRIZIONE{\TREET} of the \TREECHARFUNCTION\ \FC{\TREET} in one step and then extracts from the first record  the \OPTIMALSET\ and the expected profit corresponding to a null bonus $x=0$.
	Obviously, the complexity lies in the  function \texttt{SolveTree} called on Line \ref{Astep:solve}  (we refer to Algorithm  \ref{A:SolveTree} for a pseudo-code description of this function).
	Then,  it will be sufficient to show that Algorithm
 \ref{A:SolveTree} (\texttt{SolveTree}) computes the \TREECHARFUNCTION\ \FCA\ of a tree \TREEA\ in time $O(|\SET|^2)$.

Building the description \DESCRIZIONEA\ of \FCA\ starts with the description \DESCRIZIONE{\SUBTREE[0]{A}} of its root sub-tree \SUBTREE[0]{A} (see Section \ref{SEC:MATRIOSKA2}) and iteratively cumulates the descriptions \DESCRIZIONE{\SUBTREE[i]{A}} of all other sub-trees \SUBTREE[i]{A} for $i=1,\ldots,m$. 
As commented in Proposition \ref{P:LIMTE_MATRIOSKE} all descriptions are truncated appropriately at $\DISTA$.

\medskip
\textbf{Correctness.}
    The correctness of Algorithm  \ref{A:SolveTree} is based on Proposition \ref{P:FC_matrAlbero_contenuta_ProdottoCartesianoSottoalberi_BIS} which establishes  that the matryoshka of a group of trees can be obtained from the matryoshkas relating to a partition of the same group, and on Proposition \ref{P:cpl_algo_Merge} that guarantees this task is correctly performed by Algorithm \texttt{Merge}. 

\medskip
\textbf{Complexity.}
    To prove complexity we proceed by induction on the depth of the tree.

    If the tree \TREEA\ has zero depth (i.e. a leaf tree with no descendants), then $\TREEINDEX=\{0\}$ and the description \DESCRIZIONEA\ of its \TREECHARFUNCTION\ \FCA\ is computed in $O(1)$ time as described in Section \ref{SEC:MATRIOSKA2} from a vector of up to  $2$ records.

    Now, let us assume that the thesis holds for trees of any depth less than or equal to $h$ and show that it holds for trees of depth $h+1$. 

    At iteration $i\geq 1$, three tasks are performed:
    \begin{itemize}
        \item[(a)] description \DESCRIZIONE{\SUBTREE{A}} of \FCSA\ is computed, by the induction hypothesis, in $O(|\SUBSET{A}|^2)$ time;
        \item[(b)] description truncation up to $\DISTA$ is computed in $O(|\SUBSET{A}|)$ time;
        \item[(c)] cumulated description \DESCRIZIONEAI\ with $I=\{0,\ldots,i\}\subseteq\TREEINDEX$ is computed from \DESCRIZIONEAI[I'] with $I'=\{0,\ldots,i-1\}$ and $\DESCRIZIONEAI[\{i\}]$ in $O(|\SETAINDEXED[I]|)$ time.
    \end{itemize}%
	Since $|\SET|=\sum_{i\in\TREEINDEX}|\SUBSET{A}|$, then total time spent over all iterations is $O(|\SET|^2)$ for task (a) and $O(|\SET|)$ for task (b).
	Finally overall time spent in task (c) is $O(\sum_{i\in\TREEINDEX}|\SETAINDEXED[\{0,\ldots,i\}]|)=O(\sum_{i\in\TREEINDEX}\sum_{k=0}^{i}|\SUBSET[k]{A}|)=O(|\SET|^2)$.
	The total time is thus $O(|\SET|^2)$.
\end{proof}

\section{Conclusions}\label{SEC:conclusions}
In this paper, we analyze the probabilistic profitable tour problem on a tree and prove that it can be efficiently solved in $O(n^2)$ time where $n$ is the number of nodes. 
The problem finds application in service provision contexts
where 
customers are located on special road network typical of 
mountain areas. 

As future developments, one can consider other specific topological networks or study variants of the problem 
where budget constraints on time or costs are taken into account. 
Finally, the case of multi-valued prizes paid by the customers can be investigated.

\bibliographystyle{apalike}
\bibliography{PPTP-Tree-ArXiv}

\section*{Appendix}
\begin{proof}[Proof of Proposition \ref{P:prop_operazioni_insiemi}]
	Once we have shown that  the following expression holds
	\begin{align*}
		\MRICAVO{S_1}+\MRICAVO{S_2} & = \MRICAVO{S_1\cup S_2} + \MRICAVO{S_1\cap S_2}\\
		\MCOSTOB{S_1}{x}+\MCOSTOB{S_2}{x} & \geq \MCOSTOB{S_1\cup S_2}{x} + \MCOSTOB{S_1\cap S_2}{x}\\
	\end{align*}
it can be easily proved that:
	\begin{align*}
		\MGUADAGNOB{S_1\cup S_2}{x}  &  =\MRICAVO{S_1\cup S_2}-\MCOSTOB{S_1\cup S_2}{x}\\
		&  \geq\left[  \MRICAVO{S_1}+\MRICAVO{S_2}-\MRICAVO{S_1\cap S_2}\right]  -[\MCOSTOB{S_1}{x}+\MCOSTOB{S_2}{x}-\MCOSTOB{S_1\cap S_2}{x}]\\
		&  \geq [\MRICAVO{S_1}-\MCOSTOB{S_1}{x}]+[\MRICAVO{S_2}-\MCOSTOB{S_2}{x}]-[\MRICAVO{S_1\cap S_2}-\MCOSTOB{S_1\cap S_2}{x}]\\
		&  \geq \MGUADAGNOB{S_1}{x}+\MGUADAGNOB{S_2}{x}-\MGUADAGNOB{S_1\cap S_2}{x}.
	\end{align*}
	
The first identity can be verified immediately.
	We demonstrate the second one by induction on the depth of the tree \TREEA\ such that
	 $S_1\cup S_2\subseteq\SET$.
	The inequality is certainly true if \TREEA\ is a leaf; in this case \SET\ 
	consists of only one node and each term of the inequality is then calculated on an empty set and, is zero, or it is calculated on exactly the same node and takes the same value. It is easy to verify that the inequality holds in all cases where the leaf belongs to 
	 $S_1$ and/or $S_2$.
	
	Now, let us assume that 
	the inequality holds for any tree with depth not larger than $n$ for a fixed $n\geq 0$ and 
	show that it also holds for any tree with depth $n+1$. Let \TREEA\ be such a tree. 
	Developing the terms of inequality through \eqref{eq:costo_atteso_con_sviluppo_bonus} and \eqref{eq:costo_atteso_con_sviluppo_sottoalberi}, we get:
	\begin{align*}
		& \MCOSTOB{S_1\cup S_2}{x}=(\DISTA-x)\PSETS{(S_1\cup S_2)}+\SUMTREE\MCOSTOB{(S_1\cup S_2)\cap\SUBSET{A}}{\DISTA}\\
		& \MCOSTOB{S_1\cap S_2}{x}=(\DISTA-x)\PSETS{(S_1\cap S_2)}+\SUMTREE\MCOSTOB{S_1\cap S_2\cap\SUBSET{A}}{\DISTA}\\
		& \MCOSTOB{S_1}{x}=(\DISTA-x)\PSETS{S_1}+\SUMTREE\MCOSTOB{S_1\cap\SUBSET{A}}{\DISTA}\\
		& \MCOSTOB{S_2}{x}=(\DISTA-x)\PSETS{S_2}+\SUMTREE\MCOSTOB{S_2\cap\SUBSET{A}}{\DISTA}.\\
	\end{align*}
Thanks to induction's assumption, it is immediate to verify that:
	\[
	\SUMTREE\MCOSTOB{(S_1\cup S_2)\cap\SUBSET{A}}{\DISTA} +
	\SUMTREE\MCOSTOB{S_1\cap S_2\cap\SUBSET{A}}{\DISTA} \leq 
	\SUMTREE\MCOSTOB{S_1\cap\SUBSET{A}}{\DISTA} +
	\SUMTREE\MCOSTOB{S_2\cap\SUBSET{A}}{\DISTA}.
	\]
Then, it is sufficient to show that:  
	\[
	(\DISTA-x)\PSETS{(S_1\cup S_2)} + (\DISTA-x)\PSETS{(S_1\cap S_2)} \leq
	(\DISTA-x)\PSETS{S_1} + (\DISTA-x)\PSETS{S_2}
	\]
	or equivalently
	\[
	\PSETS{(S_1\cup S_2)} - \PSETS{S_1}  \leq
	\PSETS{S_2} -\PSETS{(S_1\cap S_2)}
	\]
	recalling that  $\PSETS{S}=1-\PNOSETS{S}$,
	we develop the two members of the inequality and we collect a common factor obtaining
	\[
	\PNOSETS{S_1}\cdot\left[1-\PNOSETS{(S_2\setminus S_1)}\right] \leq
	\PNOSETS{(S_1\cap S_2)}\cdot\left[1-\PNOSETS{(S_2\setminus S_1)}\right]
	\]
	which can be further simplified by removing the factors common to the two members, first in
	\[
	\PNOSETS{S_1} \leq
	\PNOSETS{(S_1\cap S_2)}
	\]
	and finally in 
	\[
	\PNOSETS{(S_1\setminus S_2)} \leq 1
	\]
	which concludes the proof.
\end{proof}

\bigskip
\begin{proof}[Proof of Proposition \ref{P:SOGLIE}]
We prove only the first group of equations
\eqref{eq:SOGLIA_CB_espressione}, \eqref{eq:SOGLIA_CB_monotonia_C}, \eqref{eq:SOGLIA_CB_monotonia_B}. The second group derives in a completely analogous way for symmetry.
	\begin{enumerate}
		\item Formula \eqref{eq:SOGLIA_CB_espressione}.\\
		Using formula \eqref{eq:profitto_con_virtual_bonus} of Proposition \ref{P:profitto_con_virtual_bonus} we get
		\begin{align*}
			\MGUADAGNOB{\MATSETB{i}\cup \MATSETC{h}}{x} &= \MGUADAGNOB{\MATSETB{i}}{x} + \MGUADAGNOB{\MATSETC{h}}{\VBENZ{\MATSETB{i}}}
		\end{align*}
		therefore 
		$$
		\MGUADAGNOB{\MATSETB{i}\cup \MATSETC{j}}{x}=\max_{h}\MGUADAGNOB{\MATSETB{i}\cup \MATSETC{h}}{x}
		$$
		if and only if		
		$$
		\MGUADAGNOB{\MATSETC{j}}{\VBENZ{\MATSETB{i}}}=\max_{h}\MGUADAGNOB{\MATSETC{h}}{\VBENZ{\MATSETB{i}}}
		$$
		which happens, by construction, if and only if		$$\RDXMINBC{\GRUPPOC}{j}\leq \VBENZ{\MATSETB{i}} \leq \RDXMAXBC{\GRUPPOC}{j}$$
			that is for
		$$x\in\left[\REQUIREREC{\GRUPPOC}{j}{\MATSETB{i}},\REQUIRERECMAX{\GRUPPOC}{j}{\MATSETB{i}}\right]$$
		and thus $\SOGLIA{\MATSETC{j}}{\MATSETB{i}}=\REQUIREREC{\GRUPPOC}{j}{\MATSETB{i}}$. 
		
	Observe that, being $\PSETS{\MATSETB{i}}>0$ and $\RDXMINBC{\GRUPPOC}{j}\leq\DISTA$, it holds that $\SOGLIA{\MATSETC{j}}{\MATSETB{i}}\leq\RDXMINBC{\GRUPPOC}{j}$.
		
		\item Chain \eqref{eq:SOGLIA_CB_monotonia_C}.\\
		It comes directly from \eqref{eq:SOGLIA_CB_espressione} and increasing monotony of the points \RDXMINBC{\GRUPPOC}{j} with respect to index  $j$.
		
		\item Chain \eqref{eq:SOGLIA_CB_monotonia_B}.\\
	It comes directly from \eqref{eq:SOGLIA_CB_espressione} and increasing monotony of the probabilities \RDPENDENZABC{\GRUPPOB}{i} with respect to index $i$.
	\end{enumerate}
\end{proof}

\bigskip

\begin{proof}[Proof of Proposition \ref{P:MTK_da_BiCj_espandi_matrioska_coerente}]
	The first two cases are trivial. 
	If $i<\CARDB$ and $j=\CARDC$, then all elements of \MATRIOSKAA\ not yet contained in \MATRIOSKAKH\ are of the type $\MATSETB{i+k}\cup \MATSETC{\CARDC}$. 
	Since $\MATSETB{i+1}\cup \MATSETC{\CARDC}$ is the smallest of these sets, setting  $\MATSETK{h+1}=\MATSETB{i+1}\cup \MATSETC{\CARDC}$ allows us to increase the matryoshka without compromising coherence. 
	Case $i=\CARDB, j<\CARDC$, is discussed symmetrically.

Excluding the first two cases, of the remaining two we treat only the first; the last one derives in a completely analogous way for symmetry.
	
	We therefore assume that $i<\CARDB$, $j<\CARDC$ and $\SOGLIA{\MATSETB{i+1}}{\MATSETC{j}}\leq\SOGLIA{\MATSETC{j+1}}{\MATSETB{i}}$. 
	We show that no set of the type $\MATSETB{i}\cup \MATSETC{j+k}$ with $k\geq1$ can be an \OPTIMALSET. 
	Consequently, adding  $\MATSETK{h+1}=\MATSETB{i+1}\cup \MATSETC{j}$ to \MATRIOSKAKH, not only retains the matryoshka properties for \MATRIOSKAKH[h+1], but also retains coherence with \MATRIOSKAA\ because the choice made excludes only sets proved not to belong to \MATRIOSKAA\ and $\MATSETK{h+1}$ remains the smallest set of \CARTMATRIOSKABIN\ containing $\MATSETK{h}$.
	
	We proceed by showing first that sets of the type  $\MATSETB{i}\cup \MATSETC{j+k}$ with $k\geq1$ cannot be \OPTIMALSETS\ if $x<\SOGLIA{\MATSETB{i+1}}{\MATSETC{j}}$. 
	Indeed, $x<\SOGLIA{\MATSETC{j+1}}{\MATSETB{i}}$ also holds and therefore $x<\SOGLIA{\MATSETC{j+k}}{\MATSETB{i}}$ for any $k\geq1$ due to the monotony \eqref{eq:SOGLIA_CB_monotonia_C} with respect to the entering set. It follows, for the definition of \SOGLIAINGRESSO, that $\MATSETB{i}\cup \MATSETC{j+k}$ is strictly dominated when  $x<\SOGLIA{\MATSETB{i+1}}{\MATSETC{j}}$.
	
	\medskip
	
	We now proceed to show that sets of the type $\MATSETB{i}\cup \MATSETC{j+k}$ with $k\geq1$ cannot be \OPTIMALSETS\ even if $x\geq\SOGLIA{\MATSETB{i+1}}{\MATSETC{j}}$,
	
	For the monotony \eqref{eq:SOGLIA_BC_monotonia_C} with respect to the committed set we have
	$$
	\SOGLIA{\MATSETB{i+1}}{\MATSETC{j+k}} < \SOGLIA{\MATSETB{i+1}}{\MATSETC{j}}  \text{ for all } k\geq1
	$$
	so $x>\SOGLIA{\MATSETB{i+1}}{\MATSETC{j}}$ implies $x>\SOGLIA{\MATSETB{i+1}}{\MATSETC{j+k}}$ for each $k\geq1$. Now let $i'\geq i+1$ be the maximum index for which $x>\SOGLIA{\MATSETB{i'}}{\MATSETC{j+k}}$ and, from the definition of \SOGLIAINGRESSO
	$$
	\MGUADAGNOB{\MATSETB{i}\cup \MATSETC{j+k}}{x}<\MGUADAGNOB{\MATSETB{i'}\cup \MATSETC{j+k}}{x},
	$$
	which proves the non-optimality of sets $\MATSETB{i}\cup \MATSETC{j+k}$.
\end{proof}

\end{document}